\documentclass[twocolumn,showpacs,amsfonts,amssymb,floats,aps,prb,10pt]{revtex4-1}
\usepackage{amsmath,amssymb,hyperref,url,braket,graphicx,color}
\usepackage[english]{babel}
\usepackage{amsthm}
\usepackage{amsfonts}
\usepackage{bbm}
\usepackage{mathtools}
\usepackage[utf8]{inputenc}
\usepackage{dsfont}

\newcommand\leftidx[3]{%
 {\vphantom{#2}}#1#2#3%
}

\begin{document}
\title{Increased coherence time in narrowed bath states in quantum dots}

\author{Lars B. Gravert}\email{lars.gravert@tu-dortmund.de}
\author{Peter Lorenz}
\author{Carsten Nase}
\author{Joachim Stolze}
\author{G\"otz S. Uhrig}
\affiliation{Lehrstuhl f\"{u}r Theoretische Physik I, 
Technische Universit\"{a}t Dortmund,
 Otto-Hahn Stra\ss{}e 4, 44221 Dortmund, Germany}

\begin{abstract}
We study the influence of narrowed distributions of the nuclear Overhauser field on the decoherence
of a central electron spin in quantum dots. We describe the spin dynamics in quantum dots
by the central spin model. We use analytic solutions for uniform couplings and the time
dependent density-matrix renormalization group (tDMRG) for nonuniform couplings. 
With these tools we calculate the dynamics of the central spin for large baths of nuclear spins with or without external magnetic field applied to the central spin.  The focus of our study is the influence of initial mixtures with narrowed distributions of the Overhauser field and of applied magnetic fields on the decoherence of the central spin.
\end{abstract}
\maketitle
	
\today	


% 03.65.Yz Decoherence; open systems; quantum statistical methods
% 03.65.Sq Semiclassical theories and applications
% 72.25.Rb Spin relaxation and scattering
% 75.75.-c Magnetic properties of nanostructures
% 78.67.Hc Optical properties of low-dimensional, mesoscopic, and nanoscale materials and structures
% here: 	Quantum dots
	
\section{Introduction}

In the last two decades, quantum dots have been studied intensively as realizations of 
quantum bits in theory \cite{loss98,merku02,khaet02} as well as in experiment 
\cite{kikka98,greil06a,hanso07}.
In these systems, an electron or a hole is confined in all three spatial dimensions
which explains the term ``quantum dots''. Henceforth we address the spin of
such an electron or hole as ``electronic spin'' or ``central spin''.
By suitable manipulations, this electronic spin can be controlled and prepared 
\cite{greil06b,greil07a,greil07b,xu07}. The central electronic spin and the spin bath 
given by nuclear spins of the surrounding solid are coupled by the hyperfine interaction
\cite{merku02,schli03}. Due to this coupling 
the electronic spin decoheres, hence it loses its prepared initial state within a specific
 time scale called the coherence time. Indeed, suppressing the decoherence, i.e., 
prolonging the coherence time, is one of the challenging issues in the realization of 
quantum bits (qubits) in quantum dots. For any practical use in a quantum information device 
the coherence time has to be long enough to allow a certain number of logical operations 
applied to the quantum bit.

There are various ways to suppress decoherence, i.e., to prolong the possibility
of coherent manipulations. Dynamic decoupling consists of appropriate pulse sequences which 
can increase the coherence time significantly. Theoretical 
\cite{viola98,witze07a,uhrig07,uys09} and 
experimental \cite{bierc09a,du09,barth10,bluhm10b,lange10}
studies have shown that dynamic decoupling is indeed a powerful strategy.

An alternative approach to enhance the coherence of the central spin is to polarize the 
nuclear spins coupled to the central spin \cite{imamo03,gulla10,schue14,econo14,smirn15}. 
Due to the polarization the fluctuations of the nuclear spins are reduced and
hence the central spin decoheres more slowly. 
According to Ref.\ \onlinecite{coish04}, however, very high polarizations are needed to increase
the coherence time significantly. The required large polarizations are not achieved in 
experiments so far, see Refs.\ \onlinecite{brack05,baugh07}.

A third approach to reach longer coherence consists of decreasing the 
fluctuations of the Overhauser field without polarizing the nuclear spins. 
Theoretical \cite{stepa06,klaus06,danon08,issle10,onur14} and 
experimental \cite{greil07a,latta09,vink09,xu09,bluhm10a} studies present preparation 
techniques to realize such narrowed distributions of the Overhauser field in the initial states of the quantum dots. One crucial issue is to what extent the coherence time can be increased. 

In the present paper, we will analyze the influence of 
different variances of the Overhauser field as well as the effect of an external magnetic 
field applied to the central spin. Our study is based on spin baths with about 50 to 1000 spins
which are treated either analytically for uniform couplings between the central spin
and all bath spins or numerically for exponentially distributed couplings. 

The paper is set up as follows. First, the model is presented in Sec.\ \ref{sec:model}.
Next, in Sec.\ \ref{sec:method} the two methods used are shown. In the subsequent
Sec.\ \ref{sec:results} the results are presented. Finally, the paper is concluded
in Sec.\ \ref{sec:conclusion}. Various technical aspects are explained in the appendices.

\section{Model} 
\label{sec:model}

We study the dynamics of an electron or hole spin in a quantum dot. This central spin is 
surrounded by nuclear spins. In many studies, quantum dots are composed of nuclear spins 
with spin quantum  numbers $I>1/2$ as they are present in Al, As, Ga, and In \cite{lee05,petro08}. 
Nonetheless we use $I = 1/2$ in our calculations  for numeric simplicity.

In order to describe quantum dots, we use the central spin model  (CSM)
first proposed by Gaudin \cite{gaudi76,gaudi83}. 
In this minimal model, the nuclear spins are coupled to the central spin, but not directly 
to each other. The dipolar couplings between the nuclear spins are neglected because they 
are small in comparison to the dominant hyperfine couplings \cite{merku02,schli03}. 
Hence they are not important for the time scale we analyze.

In addition to the couplings between the spins we introduce an external magnetic field. This 
field is  applied only to the central spin due to the smallness of the magnetic moment $\mu_\text{nucl}$ 
of the nuclear  spins as compared to the electronic magnetic moment $\mu_\text{B}$.

The considered Hamiltonian of the CSM has the form 
\begin{subequations}
\begin{align}
 \hat{H} &= \sum_{k = 1}^{N} A_k \hat{\mathbf{I}}_k \hat{\mathbf{S}} - h \hat{S}_z 
\\
&= \hat{\mathbf{B}} \hat{\mathbf{S}} - h \hat{S}_z 
\label{eq:Gaudin}
\end{align}
\end{subequations}
with the central spin $\hat{\mathbf{S}}$, the magnetic field $h$ in the $z$ direction, 
the $k$th bath spin $\hat{\mathbf{I}}_k$, and the corresponding coupling strength $J_k$. 
The operator  $\hat{\mathbf{B}}$ is the sum over all nuclear spins weighted 
with the corresponding coupling constant.

It is instructive to decompose the Hamiltonian according to
\begin{align}
 \hat{H} = \frac{1}{2} \left(\hat{B}^{+} \hat{S}^{-} + \hat{B}^{-} \hat{S}^{+} \right) + 
\hat{B}_z \hat{S}_z - h \hat{S}_z.
\end{align}
In this form, we can identify two parts in the Hamiltonian. The flip-flop terms 
$\hat{B}^{+} \hat{S}^{-} + \hat{B}^{-} \hat{S}^{+}$ induce spin transfer between the central spin and the nuclear spins. While the $z$ component of the total 
spin $\hat{S}_{\text{tot},z} = \sum_{k=1}^N\hat{I}_{k,z} + \hat{S}_z$ is a conserved quantity 
$\hat{{B}}_z$ and $\hat{{S}}_z$ separately are not constant in time. 
The flip-flop terms increase the decoherence of the central spin.
Thus, it is desirable to suppress them. 
In Refs.\ \onlinecite{fisch08a,teste09,hackm14a} the influence on anisotropic couplings 
of a hole spin are investigated. Due to the anisotropy the flip-flop terms are suppressed. This leads to an increased coherence time. Recently, the additional 
effect of quadrupolar terms was considered and it was shown that it is
in practice difficult to tell the various anisotropic effects apart \cite{hackm15}. 

The remaining longitudinal terms in the $z$ direction do not change the $z$ components 
but induce a Larmor precession around the $z$ axis. 
The nuclear spins induce an effective field acting on the central spin called the 
Overhauser field, denoted by $\hat{B}_z$. It leads to a shift in the effective magnetic
field applied to the central spin. Since the Overhauser field fluctuates the central 
spin dephases in its time evolution. Hence even without the flip-flop terms these 
fluctuations destroy the coherence of the central spin. This mechanism of dephasing is present 
for both hole spins and electron spins. By reducing the initial variance of the Overhauser field 
$\hat{B}_z$, i.e., by narrowing the distribution of $\hat{B}_z$, one expects to increase the 
coherence. For high magnetic fields previous papers \cite{stepa06,onur14} showed 
that the coherence time is inversely proportional 
to the width of the distribution of the Overhauser field.

\subsection{Couplings} 
\label{sec:model_couplings}

The CSM can be investigated for various distributions of the couplings $A_k$. 
To compare time scales for different sets of couplings we introduce the energy scale
\begin{align}
 A_Q \coloneqq \sqrt{\sum_{k = 1}^{N} A_k^2}.
\label{eq:CouplingsNorm}
\end{align}
All energies will be expressed in units of $A_Q$ and all times in units of $1/A_Q$
 setting $\hbar$ to unity.

Uniform couplings, i.e., $A_k = A_Q/\sqrt{N} \; \forall k$, are the most simple
assumption. We are able to derive analytic results 
for this choice of couplings on the basis of previous
derivations \cite{bortz07a,coish07}. We will use these results 
to (i) study the influence of the narrowed distributions of the 
Overhauser field for uniform couplings and to (ii) gauge the accuracy
of the numerical DMRG data.

For a more realistic description of the couplings in quantum dots 
we also consider exponentially distributed couplings \cite{coish04,farib13a,farib13b}
\begin{align}
 A_k =  \sqrt{\mathcal{N}} \exp \left( - k \frac{x}{N+1} \right) .
\label{eq:inhomogeneous_couplings}
\end{align}
This form describes the coupling constants 
for a localized electron in a Gaussian orbital ground state in a two dimensional quantum dot. 
Any other exponentially localized wave function will lead to rather similar distributions
with tails of very weakly coupled spins. Since we are interested in the generic
behavior but not in details of a particular system we focus on the distribution
\eqref{eq:inhomogeneous_couplings}.
The normalization factor $\mathcal{N}$ is related to the energy scale $A_Q$ according to 
\begin{align}
 \mathcal{N} = A_Q^2 \frac{1 - \exp \left(- 2 x/(N+1) \right)}{\exp \left(- 2 x/(N+1) \right) 
- \exp \left( - 2x \right)}.
\end{align}
The spread parameter $x$ in the exponential function determines the spread of the couplings, i.e.,
the ratio between the smallest and the largest coupling which is roughly given by $\exp(-x)$.
For $x = 0$, the spread of the couplings is zero. Thus,  we retrieve the uniform case with 
$\mathcal{N} = A_Q^2/N$.

\subsection{Narrowed spin baths}

We introduce a theoretical description for narrowed spin baths and investigate 
their effect on the dynamics of the central spin.

In previous DMRG calculations \cite{fried06,stane13a,stane14b} the
initial spin bath was described by the density matrix $\hat{\rho}_\text{b}$ 
\begin{align}
 \hat{\rho}_\text{b} = \hat{\mathds{1}}/2^N , 
\label{eq:rho_b_0}
\end{align}
where $\hat{\mathds{1}}$ is the  identity operator.
In this density matrix each state is obviously equally weighted.
This can be justified in thermal equilibrium
by the $\mu$eV energy scale of the hyperfine couplings which corresponds to fractions of a Kelvin. 
Thus, even at very low temperatures the nuclear spin bath will be
completely disordered and all states weighted equally.

Because of the increasing interest in narrowed spin baths 
realized by coherent control in experimental setups {\cite{bluhm10a}} 
our goal is to introduce a suitable method to describe them. 
We introduce the density matrix 
\begin{align}
 \hat{\rho}_\text{b}(\gamma) = \frac{1}{Z(\gamma)} 
\exp\left( - \gamma {\hat{B}_z}^2/A_Q^2 \right) 
\label{eq:rho_b}
\end{align}
for the narrowed spin baths. This density matrix provides a transparent way to describe
tuned fluctuations of the Overhauser field without introducing a finite polarization. 
For instance, Bluhm {\it et al}. detect fluctuations in their double quantum dot
which can be described by gaussian distributions consistently.
We neglect a finite polarization of $\hat{B}_z$ for two reasons. First,
we want to focus on the reduction of the fluctuations without polarization.
Second, one may shift the effect of a polarization into an externally applied static field $h$. 

We call the parameter $\gamma>0$ the narrowing factor because it controls the degree of reduction of Overhauser fluctuations.
The partition function $Z(\gamma)$ normalizes 
$\hat{\rho}_\text{b}(\gamma)$; this means
\begin{align}
 Z(\gamma) = \mathrm{tr} \left[ \exp\left( - \gamma {\hat{B}_z}^2/A_Q^2 \right) \right]. 
\label{eq:partition_func_general}
\end{align}
The disordered density matrix $\hat{\rho}_\text{b}$ in \eqref{eq:rho_b_0} is restored for 
$\gamma = 0$. 
For $\gamma \rightarrow \infty$ only those states contribute which minimize ${\hat{B}^2_z}$.
Thus we expect the variance of  $\hat{B}_z$ to vanish in this limit; further details 
of this behavior are discussed in Sec.\ \ref{sec:results}.

We assume the central spin to be prepared initially to point upwards. Then the density matrix $\hat{\rho}$ of the total CSM
is initially given by the tensor product
\begin{align}
 \hat{\rho} = \hat{\rho}_\text{b} \otimes \left(\hat{\mathds{1}}/2 + \hat{S}_z\right).
\label{eq:rho_total}
\end{align}
This assumption is the standard one. The underlying idea is that the central spin
represents a quantum bit which is prepared in a special state but loses coherence
subsequently due to the interaction with the bath.
Of course, more subtle protocols may also induce a certain entanglement
between central spin and its bath which may no longer be captured by the ansatz 
\eqref{eq:rho_total}.

\section{Methods}
\label{sec:method}

For the analysis of the spin dynamics in the CSM we use two 
approaches depending on the distribution of the coupling constants $A_k$.

For uniform couplings we are able to derive analytic solutions for arbitrary external magnetic
fields $h$ and narrowing factors $\gamma$. For large bath sizes the analytic solution has to 
be evaluated numerically. Still, we can easily deal with a large number of bath spins of the 
order of $N \approx 10^3$. Since the numeric effort increases only quadratically with the 
system size we can  in principle treat very large baths $N > 10^4$.

For exponentially distributed couplings we calculate the dynamics of the central spin 
by time dependent DMRG. This numerical method can be applied to a larger number of bath spins
than most other numeric techniques; $N$ can be as large as $10^3$, see Ref.\ \onlinecite{stane13a}.
We can choose a wide range of values for the external magnetic field $h$, the narrowing factor 
$\gamma$ and the coupling spread $x$. Since the evaluation of the analytic solution is faster 
than the DMRG calculation we will use the former method for uniform couplings at $x = 0$. Additionally, we test the
 accuracy of the DMRG code by comparing the results of the two methods for $x=0$.

The Bethe ansatz\cite{gaudi76,bortz10b} has also been used  to solve the
model analytically. The solutions, however, are restricted to highly polarized spin baths. Recent calculations based on the Bethe ansatz and Monte Carlo sampling \cite{farib13a,farib13b} are not
 restricted in that way. However the stochastic evaluation is restricted to moderately
large systems of 30 to 40 bath spins in practice.

\subsection{Analytic solutions for uniform couplings} 
\label{sec:methods_analytic}

In the uniform case the coupling constants are set to $A_k = A_Q/\sqrt{N}$ for all bath spins 
as noted in Sec.\ \ref{sec:model_couplings}. 
For this choice of coupling constants we can calculate the observables analytically as presented 
in Refs.\ \onlinecite{coish07,bortz07a,erbe10}. Here we briefly sketch the applied approach. By introducing the total spin $\hat{\mathbf{I}}$ of the bath, i.e.,  the sum over all bath spins 
$\hat{\mathbf{I}}_k$, we can rewrite the Hamiltonian as
\begin{align}
 \hat{H} = \frac{A_Q}{\sqrt{N}} \hat{\mathbf{I}} \hat{\mathbf{S}} - h \hat{S}_z.
\end{align}
The main advantage of the uniform case is the fact that we can treat all bath spins as one
effective spin $\hat{\mathbf{I}}$. This spin is characterized by its quantum numbers $j$ and $m$
which correspond in the standard way to the eigenvalues $j(j+1)$ and $m$ for $\hat{\mathbf{I}}^2$ and
of $\hat{I}_z$, respectively. Because $\hat{\mathbf{I}}$ is composed of $N$ spins $S=1/2$ 
the maximum of $j$ is $j_\text{max} = N/2$ while its minimum is $j_\text{min} = 0$ or 
$j_\text{min} = 1/2$ for an even or odd $N$, respectively. 
The quantum number $m$ ranges from $-j$ to $j$ as usual.

The Hamiltonian in \eqref{eq:Gaudin} acts on every realization of the state 
$\left|j,m \right \rangle$ in the same manner. Hence it is sufficient to treat one of these 
realizations and to multiply the result  with the degeneracy factor $\Gamma(N,j)$
which counts the number of states with given quantum numbers $j$ and $m$.
The factor $\Gamma(N,j)$ arises from the number of permutations to create the state 
$\left | j,m \right \rangle$ with $N$ spins $S=1/2$. 
For instance, there is only one way to have 
$\left | N/2,N/2 \right \rangle$, namely all $N$ spins are pointing up.

Using standard combinatorics and basic quantum mechanics we obtain the degeneracy factor
\begin{align}
\Gamma(N,j) = {\binom{N}{N/2 + j}} - {\binom{N}{N/2 + j+1}}
\end{align}
which depends only on the number $N$ and the quantum number $j$. {
For further details on the degeneracy factors, see Ref.\ \onlinecite{arecc72}.} 
Here we use the definition 
\begin{align}
 {\binom{n}{k}} = \begin{cases} \frac{n!}{k!(n-k)!} \quad &k \le n \\ \quad \; 0 \quad &\text{otherwise} \end{cases}
\end{align}
for the binomial coefficients.

In order to represent the density matrix $\hat{\rho}_\text{b}$ \eqref{eq:rho_b} in the basis 
spanned by the states labeled by the quantum numbers $j$ and $m$, we introduce the 
weight $g_{j,m}(\gamma)$. This weight includes the degeneracy factor $\Gamma(N,j)$,
 the exponential weight $\exp\left(-\gamma \hat{B}_z^2/A_Q^2\right)$,
 and the partition function $Z$ in \eqref{eq:rho_b}.
In this basis we can express the exponential weight in a particularly convenient form because 
it is the eigenbasis of $\hat{\mathbf{I}}$ and of 
the Overhauser field $\hat{B}_z$ which is equal to $\hat{I}_z$ times $A_Q/\sqrt{N}$. 
We easily determine the eigenvalues $\mu_B$ of $\hat{B}_z$ to be
\begin{align}
 \mu_B(m) = A_Q m/\sqrt{N}
\end{align}
in the basis labeled by $j$ and $m$. The eigenvalues are independent of the quantum number 
$j$ which is helpful in calculating the partition function $Z$.
We may carry out the sum over $j$ explicitly obtaining
\begin{align}
 Z_\text{a} = \sum_{m = -N/2}^{N/2} {\binom{N}{N/2 + m}} e^{ - \gamma m^2/N} 
\label{eq:partition_func_analytic}
\end{align}
where the sum over $m$ remains. In principle, one can calculate $Z$ analytically. For reasonable
 bath sizes $N$, however, we evaluate $Z$ numerically according to 
\eqref{eq:partition_func_analytic}.
The weight $g_{j,m}(\gamma)$ and the density matrix $\hat{\rho}_\text{b}(\gamma)$ can be expressed as
\begin{align}
 g_{j,m} &= \frac{\exp\left(- \gamma m^2/N\right) \Gamma(N,j)}{Z_\text{a}}, 
\label{eq:g_jm} 
\\
 \hat{\rho}_\text{b} &= \sum_{j,m} g_{j,m} \left | j,m \right \rangle \left \langle j,m \right |.
\end{align}
To obtain the total density matrix $\hat{\rho}$ at time $t=0$ one has to calculate the tensor
product of the bath density matrix $\hat{\rho}_\text{b}$ and the density matrix $\hat{\rho}_c$ of 
the central spin. Here and in the following sections we assume the central spin to be 
polarized upwards initially.

Since we are interested in the dynamics of the system we have to determine the time evolution of
the states. In the chosen basis, the Hamiltonian of the CSM is block diagonal 
and consists mainly of $2\times2$ blocks. The Hamiltonian couples the states 
$\left | \uparrow \right \rangle \otimes \left | j,m \right \rangle$ and 
$ \left | \downarrow \right \rangle \otimes \left | j,m+1 \right \rangle$ for $\left|m\right| <j$.
The red{remaining} cases $\left | \uparrow \right \rangle \otimes \left | j,j \right \rangle$ and 
$ \left | \downarrow \right \rangle \otimes \left | j,-j \right \rangle$ are eigenstates of 
$\hat{H}$. So we just have to diagonalize $2\times2$ blocks to compute the time evolution operator. The corresponding time dependent states and time dependent expectation values are presented in Appendix \ref{app:time_evolution}.

\subsection{Density matrix renormalization group}
\label{sec:methods_DMRG}

The density matrix renormalization group (DMRG) was introduced by White in 1992 
\cite{white92} as a method for efficient numerical renormalization in one-dimensional 
lattice systems. Since its introduction the DMRG has been extended to a wide range
of one-dimensional systems reviewed in Refs.\ \onlinecite{schol05} and 
\onlinecite{schol11}. In particular, it was established that DMRG is capable of
capturing time dependent phenomena as well \cite{white04a,daley04} leading to
the time dependent DMRG (tDMRG).

It was shown previously \cite{stane13a} that tDMRG can be used to very 
efficiently calculate time dependent observables in the CSM for very large spin baths. By using 
purification, see for instance Refs.\ \onlinecite{buhle00,karra12a}, we are able to 
directly calculate expectation values at infinite temperature. The traces are converted to
expectation values of a purified state in a doubled Hilbert space. We make use of the 
Trotter-Suzuki decomposition (TS decompostion) in second order to evolve this purified state in time \cite{stane13a}. 
More explicitly, we split the Hamiltonian into local operators $\hat{H}_k$ as follows
\begin{align}
 \hat{H} = \sum_{k=1}^N \left[A_k \hat{\mathbf{I}}_k \hat{\mathbf{S}} - \frac{h}{N} \hat{S}_z \right] = \sum_{k=1}^N \hat{H}_k.
\end{align}
These local operators act on one bath spin $k$ and the central spin. Since we want to evolve the system iteratively in time, we define the time evolution operator
\begin{align}
 \hat{U} \coloneqq \hat{U}(t,t+\Delta t) = \exp(-i\hat{H}\Delta t)
\end{align}
for a step $\Delta t$ in time. With the TS decomposition we obtain 
\begin{align}
 \hat{U} \approx \prod_{k=1}^N e^{-i \Delta t/2 \hat{H}_k} \prod_{k=1}^N e^{-i \Delta t/2 \hat{H}_{N+1-k}}.\label{eq:U_TSD}
\end{align}
This symmetric form of the short-time evolution operator \eqref{eq:U_TSD} is correct up to 
$\Delta t^3$ for any Hamiltonian that can be decomposed in a sum regardless
 of the vanishing of the commutator $[\hat{H}_k,\hat{H}_j]$ between different local parts \cite{hatan05}. 
To evolve the state over a finite time interval $T$, the number of necessary times
steps is $T/\Delta t$. Thus the accumulated error of the total evolution grows like
 $\Delta t^2$.

An important alternative ansatz to the TS decomposition has been introduced in Ref.\ 
\onlinecite{schmi04c}. It is based on recursively 
added Krylov vectors until no substantial error in each time step occurs.
Thus, no significant errors due to decomposition are introduced. The drawbacks of the
Krylov approach are increased computation time and additional required memory in comparison to 
the TS decomposition. In Ref.\ \onlinecite{stane13a}, both approaches were carefully compared
and good agreement between both methods was found unless very high accuracy is necessary 
Since the central spin model can be treated quite accurately by the TS decomposition 
we use it for the sake of efficiency.

To take the narrowed bath density matrix \eqref{eq:rho_b} into account we need to 
modify the previously used code to construct a suitable target state of the form
\begin{align}
 \left | \gamma \right \rangle = 
\exp\left( - \frac{\gamma}{2} {\hat{B}}_z^2/{A_Q^2} \right) 
\left | \psi \right \rangle, 
\label{eq:target_state}
\end{align}
where $\left| \psi \right \rangle$ is the purified state as defined in Ref.\ 
\onlinecite{stane13a}. 
For the purification, we add to each real spin of the system 
an auxiliary or ghost spin which is entangled with its real counterpart
in a singlet state at time $t = 0$ \cite{buhle00}. 
Hence the bath state $\left| S \right \rangle$  is initially given by
\begin{align}
 \left| S \right \rangle = \bigotimes\limits_{k = 1}^{N} \frac{1}{\sqrt{2}} \left(\left| 
\uparrow_r \downarrow_s \right \rangle - 
\left| \downarrow_r \uparrow_s \right \rangle \right)_k,
\end{align}
where $r$ denotes the real spin and $s$ denotes the corresponding auxiliary spin. 
At $t = 0$, the purified state $\left| \psi \right \rangle$ is given by the tensor product 
of the bath state $\left| S \right \rangle$ and the central spin state. Since we assume 
the central spin to be polarized upwards the purified state reads
\begin{align}
 \left| \psi \right \rangle = \left| S \right \rangle \otimes \left| \uparrow \right \rangle.
\end{align}

With the help of the narrowed state $\left| \gamma \right \rangle$ 
we can calculate expectation values of any  observable $\hat{O}$ 
with the density matrix $\hat{\rho}$ in \eqref{eq:rho_total} 
\begin{subequations}
\begin{align}
 \left \langle \hat{O} \right \rangle &= \mathrm{Tr}\left[\hat{O}\hat{\rho} \right] 
\\
&= 
\frac{\left \langle \gamma \right| \hat{O} \left | 
\gamma \right \rangle}{\left \langle \gamma \right . \left | \gamma \right \rangle} 
\label{eq:DMRG_expVal}
\end{align}
\end{subequations}
as shown in Ref.\ \onlinecite{buhle00}. 
Since one can diagonalize $\hat{B}_z$ numerically 
the exponential function in \eqref{eq:target_state} can be directly applied to the purified 
state $\left| \psi \right \rangle$. 
Some additional aspects must be considered because the operator 
$\hat{B}_z$ consists of operators of the environment block as well as of the system block. 
The details of the calculation are presented in Appendix \ref{app:construct_gamma}.

Starting from the states $\left | \psi \right \rangle$ and $\left | \gamma \right \rangle$ we 
construct the reduced density matrix 
\begin{align}
\hat{\rho}^\prime_{S} = \mathrm{tr}_E\left[w_\psi  \left | \psi \right \rangle \left \langle 
\psi \right | + w_\gamma \left | \gamma \right \rangle \left \langle \gamma \right | \right] 
\label{eq:red_density_prime}
\end{align}
by tracing out the environment $E$. For the weights $w_\alpha$ we choose 
$w_\psi=w_\gamma=1/2$. 
We sweep through the central spin system 
until the partition function $Z(\gamma)$ converges
within some tolerance, namely the absolute difference of $Z(\gamma)$ between two consecutive sweeps is below $10^{-8}$. However, the absolute difference is below $10^{-11}$ after the second sweep in typical cases. This partition function is easily accessed by evaluating 
the scalar product
\begin{align}
 Z_\text{n}(\gamma) = \left \langle \gamma \right. \left| \gamma \right \rangle. 
\label{eq:partition_func_DMRG}
\end{align}

We observe that by adding the density matrices 
\begin{subequations}
\begin{align}
 \hat{\rho}_1 &= \mathrm{tr}_E\left[ \hat{B}_z \left | \psi \right \rangle \left 
\langle \psi \right | \hat{B}_z \right], 
\\
 \hat{\rho}_2 &= \mathrm{tr}_E\left[ \hat{B}_z \left | \gamma \right \rangle 
\left \langle \gamma \right | \hat{B}_z \right ] 
\end{align}
\end{subequations}
to the reduced density matrix $\hat{\rho}^\prime_{S}$ the numeric accuracy can be increased
considerably. Therefore, we use the total reduced density matrix 
\begin{align}
\hat{\rho}_{S} = w_0 \hat{\rho}^\prime_S + w_1 \hat{\rho}_1 + w_2 \hat{\rho}_2,
\end{align}
with the normalized weights $w_0+w_1+w_2=1$. In Appendix \ref{app:weight} we include an 
analysis {to clarify} how the weights $w_1$ and $w_2$ influence the accuracy of 
the DMRG data. We 
find that even small weights $w_1$ and $w_2$ increase the accuracy noticeably. 
Hence we choose $w_1=w_2=1/22$ in the construction of the narrowed state 
$\left | \gamma \right \rangle$.

For the tDMRG $\left | \gamma \right \rangle$ and an additional state 
\begin{align}
 \left | \phi \right \rangle \coloneqq \hat{S}_{-} \left | \gamma \right \rangle \label{eq:phi}.
\end{align}
are the target states. This specific choice of \eqref{eq:phi} is due to the correlation function defined by
\begin{align}
C(t) =  \left \langle \hat{S}_+(t) \hat{S}_-(0) \right \rangle.
\end{align}
We elaborate on this correlation function in Sec.\ \ref{sec:decoherence}. Since $\hat{S}_{-}$ acts only upon the central spin we do not need to use a complete half-sweep to apply $\hat{S}_{-}$ to $\left | \gamma \right \rangle$ as discussed in Ref.\ \onlinecite{white04a}. Instead we can apply $\hat{S}_{-}$ once before starting the evolution in time. We can calculate the correlation function by evaluating 
\begin{align}
 C_\text{n}(t) = \left \langle \tilde{\gamma} \left| \hat{S}_{+}(t) \right| \tilde{\phi} \right \rangle
 \end{align}
with the normalized states $\left | \tilde{\gamma} \right \rangle$ and $\left | \tilde{\phi} \right \rangle$. By this construction we can calculate correlation functions very efficiently \cite{white04a}.

In the following we use $m = 1024$ states in all calculations performed with the DMRG and the tDMRG. The step size is chosen as $\Delta t = 0.01 A_Q$ in all calculations performed with the tDMRG.
	
\section{Results}
\label{sec:results}

With the help of the presented methods we are able to analyze the dynamics of the central 
spin and of the spin bath. We investigate the influence of the initial variance of 
the Overhauser field and of the external magnetic field applied to the
central spin on the dynamics of the central spin. We deal with uniform
and exponentially distributed couplings, see \eqref{eq:inhomogeneous_couplings}, based on the 
analytic solution and on the time dependent DMRG with $m = 1024$ states, respectively.

We will analyze the dependence of the variance of the Overhauser field and of the coherence 
time on the external magnetic field $h$, on the narrowing factor $\gamma$, on the spread parameter 
$x$, and on the bath size $N$. While all these parameters influence the initial variance and 
the coherence time to some degree we focus on variations in $h$ and $\gamma$ in particular 
because we observe the strongest dependences for them.

\subsection{Accuracy check of DMRG}

Before we turn to the variance and the dynamics we want to use 
the analytic results in the uniform case to determine the accuracy of the DMRG approach. We 
calculate the difference between the analytic results and the DMRG results for the 
partition function $Z(\gamma,N)$. Since this is a static quantity the 
difference depends only on the narrowing factor $\gamma$ and the number of bath spins $N$, 
but not on the external magnetic field $h$. To analyze the relative error 
\begin{align}
 \Delta Z (\gamma,N) = \left | 
\frac{ Z_\text{n}(\gamma,N)}{Z_\text{a}(\gamma,N)}  - 1 \right |
\end{align}
of the DMRG calculations with $m = 1024$ states we vary one parameter keeping the other constant. 
For the calculation of the analytic partition function $Z_\text{a}$ and of the
numeric partition function $Z_\text{n}$  we use \eqref{eq:partition_func_analytic} and 
\eqref{eq:partition_func_DMRG}, respectively. Hence the error $\Delta Z(\gamma,N)$ 
measures how much the results of the DMRG calculation differs from the analytic results.

In Appendix \ref{app:accuracy} we present in more detail that the DMRG code is capable 
of describing the narrowed states for uniform couplings very well. In summary, the 
error $\Delta Z$ is below $10^{-11}$ for nearly all parameter values we considered and of the 
order of $10^{-9}$ in the worst cases. This proves the high accuracy of the DMRG code 
in constructing the narrowed states for $x=0$, i.e., for uniform couplings. For nonuniform couplings the calculation becomes less accurate indicated by an increased discarded weight.
Nonetheless, we are still able to calculate the narrowed states reliably as shown in the next paragraph.

\begin{figure}[htb]
  \begin{center}
  \includegraphics[width=\columnwidth]{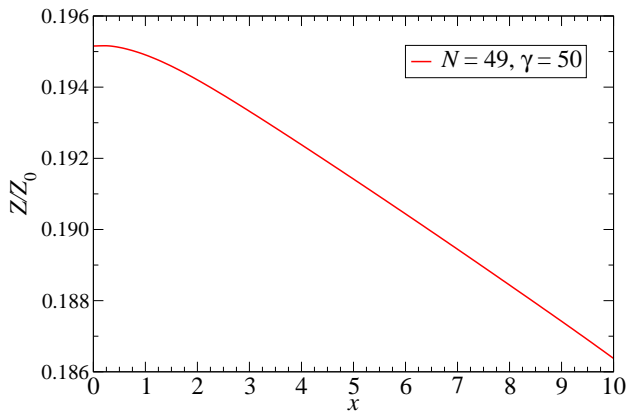}  
  \caption{The relative partition function $Z/Z_0$ for $N = 49$ and $\gamma = 50$ depending on the 
	  spread parameter $x$ in \eqref{eq:inhomogeneous_couplings}. While the spread of the couplings 
	  changes considerably according to the exponential distribution in \eqref{eq:inhomogeneous_couplings}
	  the relative partition function drops by about $5\%$ only.}
 \end{center} 
  \label{fig:Zx}
\end{figure}

To study the dependence of $Z$ on the spread parameter $x$ we plot the numeric
results $Z(x,\gamma)/Z_0(x)$ in Fig.\ \ref{fig:Zx} for a fixed narrowing factor 
$\gamma = 50$. Here $Z_0$ denotes the partition function for $\gamma = 0$. Hence 
the plotted quantity shows the relative reduction of the partition function induced by 
the narrowing. The data shows that $Z(x,\gamma)/Z_0(x)$ depends hardly
on the coupling spread $x$. We emphasize
that the ratio between the smallest and the largest
coupling in (4) is roughly given by $\exp(-x)$. Hence the
case $x = 10$ captures more than four decades of coupling strengths. 
Nonetheless, $Z(x,\gamma)/Z_0(x)$  drops by about $5\%$ only.
For other values of $\gamma > 0$, the qualitative behavior is the same.
We will also illustrate that larger values of the spread parameter $x$ do not 
influence the initial variance of the Overhauser field strongly.

We conclude that we are able to construct the desired narrowed spin bath for both uniform
and exponentially distributed couplings. With the DMRG 
approach, we can calculate various expectation values for nonuniform couplings for 
a wide range of the spread parameter $x$.

\subsection{Initial variance of the Overhauser field}

The narrowed density matrix of the spin bath in \eqref{eq:rho_b} leads to a reduced initial 
variance of the Overhauser field $\hat{B}_z$. The effect depends on the narrowing factor $\gamma$. 
To characterize the narrowed states we consider the variance $\sigma^2$ instead of $\gamma$. 
While these two values are connected by a one-to-one mapping as illustrated later in this section, 
$\sigma^2$ is a physical property of the bath while the parameter $\gamma$ is a theoretical tool to tune the former. In some cases, however, it will be more convenient to use $\gamma$ explicitly.

Generally, the variance $\sigma^2$ is defined by
\begin{align}
 \sigma^2(t) = \left \langle \hat{B}_z(t)^2 \right \rangle - \left \langle \hat{B}_z(t) 
\right \rangle^2 
\label{eq:sigma_result}
\end{align}
so that it depends on the bath size $N$, the spread parameter $x$, the narrowing factor 
$\gamma$, and the magnetic field $h$. Rigorously, the variance $\sigma^2$ is time dependent
because the Overhauser field is not a conserved quantity. 
We focus on the initial variance $\sigma^2$ at $t = 0$. Otherwise, the time dependence 
will be denoted explicitly by $\sigma^2(t)$.

The initial variance $\sigma^2$ is independent of the magnetic field $h$ because the field is 
only applied to the central spin. Hence the field $h$ influences only the dynamics of 
$\sigma^2(t)$. The expectation value $ \left \langle \hat{B}_z(t) \right \rangle$ vanishes 
at $t=0$, so that we have the simplified initial variance
\begin{align}
 \sigma^2 = \left \langle \hat{B}_z^2 \right \rangle. 
\label{eq:sigma_initial}
\end{align}
We study the dependence of the initial variance on the narrowing factor $\gamma$, the 
bath size $N$, and the spread parameter $x$. First, we discuss how 
$\sigma^2$ can be computed for uniform and nonuniform couplings.

By DMRG we calculate the variance $\sigma^2$ of the Overhauser field very fast 
and efficiently using \eqref{eq:DMRG_expVal}. We obtain
\begin{align}
 \sigma^2_\text{n} = \frac{\left \langle \gamma \right| \hat{B}_z^2 
\left | \gamma \right \rangle}{\left \langle \gamma \right . 
\left | \gamma \right \rangle}. 
\label{eq:sigma_DMRG}
\end{align}
The subscript `n' denotes solutions calculated numerically, i.e., by DMRG. In contrast to most 
other methods rather large bath sizes $N$ can be reached for nonuniform couplings $A_k$ in 
\eqref{eq:inhomogeneous_couplings}.

For uniform couplings we derive an analytic formula by calculating the derivative of 
the partition function $Z$. The relation
\begin{align}
\sigma^2 = - A_Q^2 \partial_\gamma \ln(Z) 
\label{eq:sigma_derivative}
\end{align}
holds true for each spread parameter $x$. This can be easily concluded from the 
partition function $Z$ in \eqref{eq:partition_func_general}.

Since we have an analytic expression for $Z$ in \eqref{eq:partition_func_analytic} for 
uniform couplings the variance is obtained as
\begin{align}
 \sigma^2_\text{a} = A_Q^2 \sum_{m = -N/2}^{N/2} {\binom{N}{N/2 + m}} m^2 
e^{ - \gamma \frac{m^2}{N}}. 
\label{eq:sigma_analytic}
\end{align}
The subscript `a' denotes solutions calculated by this analytic formula. Since the evaluation
effort increases linearly with $N$ very large bath sizes $N > 10^4$ can be treated in this way.

Finally, we discuss the thermodynamic limit $N\rightarrow \infty$. 
In this limit, we are able to derive the variance $\sigma^2_\infty(\gamma)$ analytically
by virtue of the central limit theorem. 
We obtain
\begin{align}
 \sigma^2_\infty(\gamma) = \frac{A_Q^2}{4 + 2\gamma} 
\label{eq:sigma_approx}
\end{align}
as shown in Appendix \ref{app:continuum_limit}.
With increasing bath size $N$, the variance $\sigma^2(\gamma)$ approaches the limit 
$\sigma^2_\infty$. Since the limit does not depend on the distribution of 
coupling constants $A_k$ the 
variance $\sigma^2_\infty(\gamma)$ is valid for uniform couplings as well as 
for nonuniform couplings. But the bath size $N$ for which the variance $\sigma^2(\gamma)$ 
can be approximated reliably by the limit $\sigma^2_\infty(\gamma)$ depends on the actual 
distribution of the coupling constants as we show here.

\begin{figure}[htb]
 \centering
 \includegraphics[width=\columnwidth]{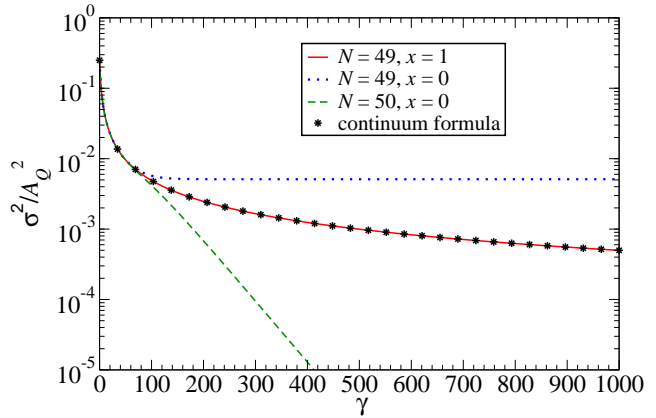}
 \caption{Variance $\sigma^2 (\gamma)$ for three sets of parameters. 
The red {solid} curve is calculated for a nonuniform system with $x = 1$ and $N=49$. 
It is approximated 
well by the thermodynamic limit \eqref{eq:sigma_approx} (black dashed curve) in
the whole interval shown. {A curve for $x = 1$ and $N=50$ is not shown because
it could not be distinguished from the $x = 1$ and $N=49$ case.}
The blue curve is calculated for an uniform system ($x = 0$) 
and $N = 49$. For large values of $\gamma$ the curve saturates to the value 
$A_Q^2/196$ according to \eqref{eq:sigma_approx_homogeneous}. The green curve is calculated 
for a uniform system with $x = 0$ and $N=50$. The curve decays as
$\exp(-\gamma/N)$ for large values of $\gamma$ 
according to \eqref{eq:sigma_approx_homogeneous}.}
 \label{fig:sigma2}
\end{figure}

First, we investigate the influence of the narrowing factor $\gamma$ on the variance. 
In Fig.\ \ref{fig:sigma2} we show three variances $\sigma^2(\gamma)$ depending on 
$\gamma$. Two curves represent the variances for uniform couplings for $N = 49$ and 
$N = 50$ bath spins, respectively. The third curve represents the variance for $N = 49$ 
bath spins and the spread parameter $x = 1$ in \eqref{eq:inhomogeneous_couplings}. 
In addition, we plot the variance $\sigma^2_\infty(\gamma)$ of the thermodynamic limit in 
\eqref{eq:sigma_approx} depicted by the dashed black line.

All curves start at the same value at $\gamma = 0$. The 
initial value $\sigma^2$ is computed analytically for every spread parameter $x$. 
By rearranging the expectation value in  \eqref{eq:sigma_initial} we obtain
\begin{subequations}
\begin{align}
\sigma^2 &= \sum_{k,l} A_k A_l \left \langle \hat{I}_{z,k} \hat{I}_{z,l} \right \rangle 
\\
 &= \sum_{k \ne l} A_k A_l \left \langle \hat{I}_{z,k} \hat{I}_{z,l} \right \rangle + A_Q^2/4.
\end{align} 
\end{subequations}
For $\gamma = 0$ the expectation value 
$\left \langle \hat{I}_{z,k} \hat{I}_{z,l} \right \rangle$ for $k\ne l$ vanishes because the 
spin operators are traceless and the density matrix of the bath spins $\hat{\rho}_\text{b}(\gamma=0)$ 
is proportional to the identity matrix. Finally, we arrive at
\begin{align}
 \sigma^2(\gamma = 0) = A_Q^2/4 \eqqcolon \sigma^2_0 
\label{eq:variance_zero}.
\end{align}
This result for $\gamma=0$ matches with the variance $\sigma^2_\infty$ of the 
continuum limit in \eqref{eq:sigma_approx}.

For increasing values of $\gamma$ the variance $\sigma^2$ decreases as shown in Fig.\ 
\ref{fig:sigma2} because states with larger values for the $z$ component of the Overhauser 
field are suppressed more and more. The fluctuations 
of the Overhauser field are a source of dephasing of the central spin. We show in 
the next subsection that this dephasing is suppressed as well.
For not too large values of $\gamma$ up to roughly 60 the variances
decrease  in all three cases as described by the approximate expression
\eqref{eq:sigma_approx}. But for even larger values of $\gamma$ the three variances start to deviate from one another.

The most obvious feature in Fig.\ \ref{fig:sigma2} is the dependence of the variances for 
uniform couplings on the parity of the number $N$ of bath spins. To analyze 
this dependence we investigate the behavior for large values of $\gamma$ 
because in this limit the differences become most pronounced.

For large values of the narrowing factor $\gamma$ the main contribution to the density matrix
$\hat{\rho}_\text{b}$ arises from the states with the lowest moduli of eigenvalues 
$|\mu_B|$ of the Overhauser field $\hat{B}_z$. 
In the uniform case $\hat{B}_z$ is proportional to the $z$ component of the momentum
of the bath $\hat{I}_z$.  Thus, the eigenvalues $\mu_B$ are proportional to the eigenvalues $m$ of 
$\hat{I}_z$. For an odd number $N$ the eigenvalue of $m$ with the lowest modulus 
is $\pm 1/2$ while it is zero for an even number $N$. 
This difference in the lowest eigenvalues 
is the source of the dependence on the parity of $N$.

For further analysis we approximate the partition function $Z(\gamma)$ in 
\eqref{eq:partition_func_analytic} for large values of $\gamma/N$ as
\begin{align}
 Z(\gamma) \approx \begin{cases} 2 {\binom{N}{N/2+1/2}} e^{-\gamma/4N} & N \text{ odd} 
\\ 
{\binom{N}{N/2}} \left(1 + 2\frac{N}{N+2} e^{-\gamma/N} \right) & N \text{ even} \end{cases}
\end{align}
by taking only the leading order into account.
Using \eqref{eq:sigma_derivative} we obtain
\begin{align}
 \sigma^2(\gamma) \approx \begin{cases} A_Q^2/(4N) & N \text{ odd} 
\\ 
\frac{2A_Q^2}{N+2} e^{-\gamma/N} & N \text{ even} \end{cases} 
\label{eq:sigma_approx_homogeneous}
\end{align}
for large values of the ratio $\gamma/N$. Any flip of a spin changes the eigenvalue $m$ by $1$.
Since the weight for the narrowed states in \eqref{eq:partition_func_analytic} is proportional to 
$\exp\left(- m^2 \gamma/N \right)$ any spin flip is exponentially suppressed.

For odd $N$ the variance $\sigma^2$ saturates at the finite value $A_Q^2/(4N)$. 
Hence we are not able to decrease the fluctuations of the Overhauser field further. In 
contrast, the variance $\sigma^2$ decreases exponentially for even $N$. 
Thus, we can  arbitrarily narrow the initial distribution of $\hat{B}_z$ in principle. 
But we consider the limit of infinite ratio $\gamma/N$ to be unphysical 
because in this limit any deviation from {uniform couplings} 
 comes more and more severely into effect.

For nonuniform couplings the situation is more complex. In this case, the Overhauser field 
$\hat{B}_z$ is not proportional to $\hat{I}_z$. Spin flips of weakly coupled spins, i.e.,
spins with {larger index $k$} in \eqref{eq:inhomogeneous_couplings}, influence the eigenvalue 
$\mu_B$ only weakly. Hence the suppression of states with larger values of $\mu_B$ 
is smoother  than in the uniform case.

\begin{figure}[htb]
 \centering
 \includegraphics[width=\columnwidth]{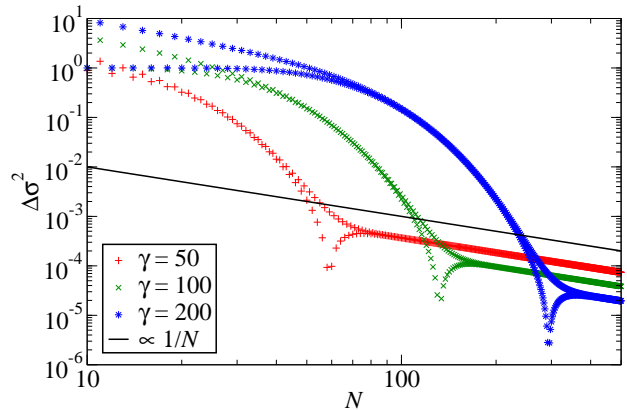}
 \caption{Relative deviation $\Delta \sigma^2$ defined in \eqref{eq:sigma_relDiff} 
depending on $N$ for three different values of the narrowing factor $\gamma$ and 
uniform couplings ($x = 0$). The black line depicts the power law $\propto 1/N$;
note the {doubly} logarithmic scale.} 
 \label{fig:sigma_N_x0}
\end{figure}

Increasing the bath size $N$ leads generally to a better agreement between 
calculations for finite spin baths and for the thermodynamic limit $N\rightarrow \infty$. 
To analyze this behavior quantitatively we study the relative deviation 
\begin{align}
  \Delta \sigma^2(N,\gamma) = \left |\frac{\sigma^2(N,\gamma)}{\sigma^2_\infty(\gamma)} 
	- 	1 \right |. 
	\label{eq:sigma_relDiff}
\end{align}
The deviation $\Delta \sigma^2(N,\gamma)$ measures how much the variance $\sigma^2(N,\gamma)$ 
for a finite bath size $N$ deviates from the corresponding variance $\sigma^2_\infty(\gamma)$ 
in the thermodynamic limit. From the results in Fig.\ \ref{fig:sigma2} we expect that 
the deviations are significantly stronger for uniform couplings than for exponentially
distributed couplings.
In Fig.\ \ref{fig:sigma_N_x0}, we plot  the relative deviation for three values of $\gamma$ 
depending on the bath size $N$ for uniform couplings.

For small bath sizes $N < 50$ all three curves show large relative deviations, especially 
for $\gamma \ge 100$. In addition, we observe jumps in $\Delta \sigma^2$ for consecutive 
values of $N$ resulting from  the dependence on the parity of the bath size $N$. These 
observations support the analysis of the data displayed in Fig.\ \ref{fig:sigma2}.
With increasing bath size $N$ the relative deviations decrease. In an intermediate range of 
bath sizes $\Delta \sigma^2$ no longer shows a strong dependence on the parity of $N$. The
position of this range differs depending on the value of $\gamma$. 
Lower values of $\gamma$ push this range to lower bath sizes $N$.
For sufficiently large bath sizes $N$ the deviation $\Delta \sigma^2$ decreases approximately 
like $1/N$ for all three values of $\gamma$.

In Fig.\ \ref{fig:sigma_N_x0}, we observe a dip like feature in $\Delta \sigma^2$ {in
the baths with even number of spins. This dip is accompanied by a 
cusp like change  of the slope for even and odd $N$. }
This feature appears for all values of $\gamma$. 
The position $N_\text{dip}$ of the dip depends on $\gamma$. 
{The three curves in Fig.\ \ref{fig:sigma_N_x0} suggest a relation $N_c\propto \gamma$.} 
The dip signals a crossing 
of $\sigma^2(N,\gamma)$ and  $\sigma^2_\infty(\gamma)$ so that $\Delta \sigma^2(N,\gamma)$ 
becomes very {small, but since the bath size $N$ is discrete, it does 
not vanish completely.} Still one can define a characteristic {even} $N_\text{c}$ 
where the crossing takes place. {For any finite odd bath size $N_\text{odd}$ 
the variance $\sigma^2(N_\text{odd},\gamma)$ lies higher than $\sigma^2_\infty(\gamma)$.}
In addition, one can define a crossing value $\gamma_\text{c}$ for any given {even} bath size $N$  fulfilling
\begin{align}
\Delta \sigma^2(N,\gamma_\text{c}) = 0.
\end{align}

We call the corresponding variance the crossing variance and denote it by $\sigma^2_c(N)$. 
Since the  deviation $\Delta \sigma^2$ vanishes for $\gamma_\text{c}$ 
we find for the crossing  variance
\begin{align}
\sigma^2_c(N) = \frac{A_Q^2}{4 + 2\gamma_\text{c}} = \sigma^2_\infty(\gamma_\text{c}). 
\label{eq:sigma_crossing}
\end{align}

Inspecting Fig.\ \ref{fig:sigma_N_x0} we see that the {bath size $N_c$ of the crossing variance also indicates the lowest bath size} above which the variance of the Overhauser field 
does no longer display sizable finite size effects. In other words, for systems larger
than $N_\text{c}$ the approximate formula \eqref{eq:sigma_approx} works well. 

\begin{figure}[htb]
 \centering
 \includegraphics[width=\columnwidth]{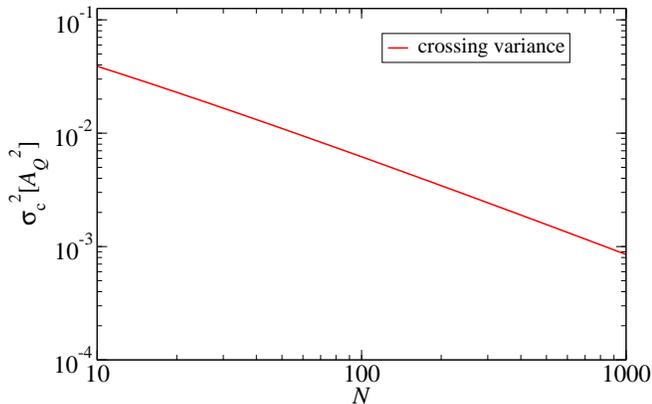}
 \caption{Crossing variance $\sigma^2_\text{c}$ defined in \eqref{eq:sigma_crossing} depending on 
the bath size $N$ for uniform couplings. This variance  represents the lowest 
variance for which the uniform system does not show significant finite 
size effects.} 
 \label{fig:sigma_crossing}
\end{figure}

In Fig.\ \ref{fig:sigma_crossing} we plot the crossing variance for uniform systems. 
From this figure, we conclude that large bath sizes $N$ are required to study 
narrowed spin baths.

\begin{figure}[htb]
 \centering
 \includegraphics[width=\columnwidth]{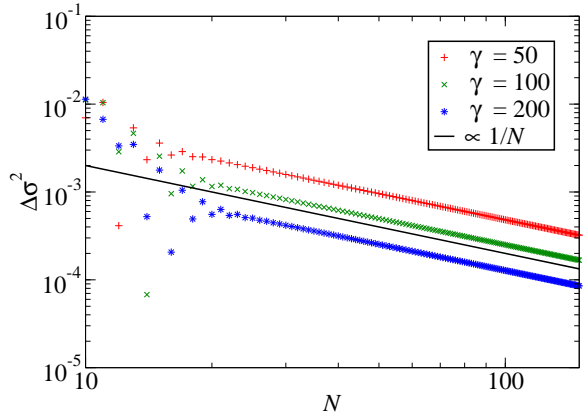}
 \caption{Relative deviation $\Delta \sigma^2$ in \eqref{eq:sigma_relDiff} depending on 
$N$ for three different values of the weight factor $\gamma$ and nonuniform couplings 
with the spread parameter $x = 1$. The black line is a power law $\propto 1/N$.}
 \label{fig:sigma_N_x1}
\end{figure}

In the case of nonuniform couplings with spread parameter $x=1$ the deviation 
$\Delta \sigma^2$ behaves quite differently. In Fig.\ \ref{fig:sigma_N_x1} we plot three 
curves for the same values of $\gamma$ as in Fig.\ \ref{fig:sigma_N_x0}. {Note that} 
the curves  start at values two orders of magnitude lower than in the uniform case. 
For small bath sizes $N<30$, 
we still observe some dependence of $\Delta \sigma^2$ on the parity of the $N$. Then, however,
the curves quickly follow the power law $\propto1/N$, i.e., they become independent of 
the parity of $N$. These results corroborate our above argument that nonuniform couplings 
dampen the finite size effects of the bath.

We conclude that for uniform and for nonuniform couplings we can find sufficiently
large baths $N>N_\text{c}$ such that the behavior of the system depends hardly on $N$. 
Hence we do not need to investigate the influence of $N$ on the coherence
time in the next subsection because we choose $N$ large enough to {observe}
the thermodynamic limit essentially.

\begin{figure}[htb]
 \centering
 \includegraphics[width=\columnwidth]{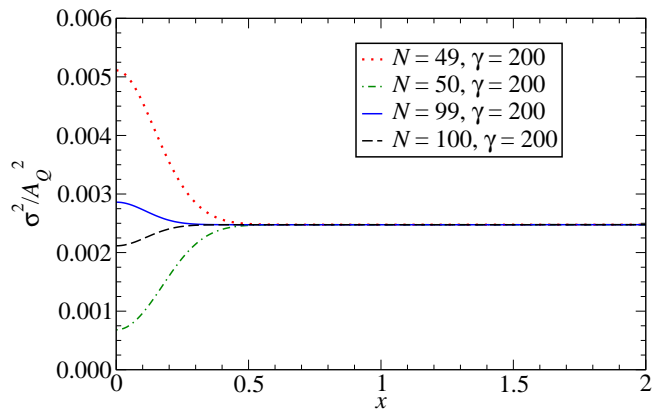}
 \caption{Variance $\sigma^2$ for the narrowing factor $\gamma = 200$ and four bath 
sizes $N$ depending on the spread parameter $x$. Both curves quickly reach the 
thermodynamic limit $\sigma^2_\infty(\gamma~=~200)~=~A_Q^2/404$ according to 
\eqref{eq:sigma_approx}.} 
 \label{fig:sigma_x}
 \end{figure}

Finally, we study the influence of the spread parameter $x$ on the variance. 
In Fig.\  \ref{fig:sigma_x} the variance $\sigma^2$ and its dependence on the coupling 
parameter $x$ is depicted.
We have chosen four bath sizes $N = 49$, $N = 50$, $N = 99$, and $N = 100$ to capture a possible dependence 
of $\sigma^2$ on the parity $N$ as discussed before. The narrowing factor is 
$\gamma = 200$. The variances $\sigma^2$ quickly approach the thermodynamic variance 
$\sigma^2_\infty(\gamma=200) = A_Q^2/404$. For spread parameters
$x>0.6$, there is no visible difference between the variances for $N=49$ and $N=50$. For larger bath sizes $N=99$ and $N=100$ the variances converge faster to the thermodynamic variance. For $x>0.4$ the corresponding curves already show no visible difference. For even larger bath sizes the variances will converge even faster so that for $N > N_\text{c}$ even for $x=0$ we are still able to capture the physics of the thermodynamic variance if $\gamma<\gamma_\text{c}$. In this limit the variance is indeed almost independent of the spread parameter $x$. 

Due to the saturation the spread parameter $x$ influences the variance $\sigma^2$ only weakly 
once we reach a certain threshold. The exact value depends on the bath size $N$ and the 
narrowing factor $\gamma$. For sufficiently large bath sizes even uniform couplings can be chosen
to analyze the coherence time as discussed before in this section. To calculate the coherence time we will choose a large bath size $N$ and uniform couplings, i.e., $x=0$ as well as a smaller bath size $N$ and nonuniform couplings with a spread parameter of $x=1$.

\subsection{Decoherence} \label{sec:decoherence}

The central spin is prepared initially to be polarized upwards. Due to the 
interaction in the Hamiltonian \eqref{eq:Gaudin} the central spin decoheres 
in the course of its temporal evolution. We call the characteristic time scale
for which the central spin keeps its initial state the ``coherence time''.
One of the crucial goals of quantum information processing is to maintain coherence
as long as possible. Thus, we aim at long coherence times.

To analyze the dependence of the coherence time on the initial variance $\sigma^2$ in 
\eqref{eq:sigma_initial} of the Overhauser field $\hat{B}_z$ and the external magnetic 
field $h$ in the Hamiltonian \eqref{eq:Gaudin} we introduce the correlation 
function
\begin{align}
 C(t) =  \left \langle \hat{S}_+(t) \hat{S}_-(0) \right \rangle. 
\label{eq:correlation}
\end{align}
Since the central spin is fully polarized upwards for $t = 0$ one has $C(0) = 1$.
In its temporal evolution the modulus $C(t)$ decreases and decays towards zero. 
By narrowing the variance $\sigma^2$ this decay is slowed down as shown in this
section.

We define the coherence time $T_2$ by the relation
\begin{align}
 \left| C(T_2) \right| = e^{-1}. 
\label{eq:T2}
\end{align}
For nonuniform couplings, we calculate $C(t)$ for equidistant time steps iteratively by
tDMRG and determine $T_2$ by linear interpolation between these time steps.
For uniform couplings we can evaluate \eqref{eq:correlation} at arbitrary time $t$, see \eqref{eq:correlation_analyticA} and \eqref{eq:correlation_analyticB}. Hence 
root-finding methods can be used to find the instant fulfilling \eqref{eq:T2}.
Note that the key idea of using the correlation \eqref{eq:correlation} is to eliminate the
main effect of Larmor oscillations about the $z$ axis.

While $C(t)$ decreases in time, it still shows some oscillations remaining from
the Larmor precession depending on the external magnetic field $h$. 
Since we define the coherence time by a threshold it may happen that 
$T_2$ jumps for particular fields from one maximum of the oscillation to the next. 
Increasing $h$ shifts the maximum to lower times so that the coherence time 
decreases until it jumps to the next maximum. Due to this behavior the graphs 
$T_2(\sigma^2)$ and $T_2(h)$ display sawtooth like features which are superposed to the
overall trend of decay. The sawtooth behavior is an artifact of 
our way to determine $T_2$; but it does not conceal the overall behavior.

For  high magnetic fields we can derive an analytic solution for the correlation 
function $C(t)$ in the thermodynamic limit $N\rightarrow \infty$. 
Neglecting the flip-flop  terms yields
\begin{align}
 C_\infty(t) = \exp\left(-\frac{\sigma^2_\infty t^2}{2}+\mathfrak{i} ht\right) 
\label{eq:highfield_limit}
\end{align}
for the correlation function with the variance $\sigma^2_\infty$ in 
\eqref{eq:sigma_approx}. In Appendix \ref{app:high_field_limit} we present 
the calculation in detail. The coherence time in this limit reads
\begin{align}
T_{2,\infty} = \sqrt{2}/\sigma_\infty 
\label{eq:T2_infty}.
\end{align}
The coherence time $T_{2,\infty}$ in the high-field limit can be increased arbitrarily by 
reducing the initial variance $\sigma^2_\infty(\gamma)$. Without narrowing, 
i.e., for $\sigma^2 = A_Q^2/4$, our result is consistent with previous papers. 
For instance in 
Ref.\ \onlinecite{cywin11}, Eq.\ $(10)$ describes the same coherence time; 
it is denoted by $T_2^*$ there. In Ref.\ \onlinecite{onur14} the coherence time $T_2$ is 
calculated in the presence of a narrowed Overhauser field and strong magnetic field 
with the same result as ours.

In the subsections below, we analyze the effects of 
the initial variance $\sigma^2$ and of the external magnetic field 
$h$ on the coherence time $T_2$. In the previous
subsection, we showed that the spread parameter $x$ and the bath size $N$ hardly influence 
the variance $\sigma^2$ for suitable ranges of parameters, i.e., $N>N_\text{c}$ and $x$
not too small. Hence the coherence time 
$T_2$ is nearly independent of these parameters as well; it changes only by a few 
percent at most. Thus we restrict ourselves to two 
representative sets of parameters: (i) exponentially distributed couplings with $x = 1$
 and $N = 49$ and (ii) uniform couplings and $N = 999$ bath spins. 
For these parameter sets the coherence time $T_2$ 
is mainly influenced by the initial variance $\sigma^2$ and the magnetic field $h$.

Since the truncation error of the tDMRG grows over time, see Ref.\ \onlinecite{stane13a}, 
we need to decide up to which truncation error we can consider the calculated data
reliable. The truncation error arises from the discarded weight in the DMRG steps,
and it grows exponentially in time \cite{gober05}.
It is not due to the Trotter-Suzuki  decomposition which constitutes also 
a source of a systematic error, but is much better controllable \cite{stane13a}.

To quantify the truncation error of the tDMRG we choose the accumulated discarded weight
\begin{align}
 \varepsilon_\text{acc} = \sum_j \varepsilon_j = \sum_j (1 - \sum_{n} w_{j,n}),
\end{align}
which is the sum of all discarded weights $\varepsilon_j$ of the $j$th step in the tDMRG\cite{stane13a}. Another suitable measure is the accumulated discarded entropy\cite{braun14} 
\begin{align}
 S_\text{trunc} = \sum_j S_{\text{trunc},j} = \sum_j \left(-\sum_n w_{j,n} \log(w_{j,n})\right),
\end{align}
which is the sum of all discarded entropies $S_{\text{trunc},j}$ of the $j$th step in the tDMRG. In our calculations $S_\text{trunc}$ behaves qualitatively very similar
 to $-\varepsilon_\text{acc} \log(\varepsilon_\text{acc})$. 
Hence we finally choose $\varepsilon_\text{acc}$ for numerical simplicity.
 
For the accumulated discarded weight $\varepsilon_\text{acc}$ we choose the threshold error 
$\varepsilon_\text{th} = 10^{-3}$. If $\varepsilon_\text{acc}$ exceeds this value
we do not push the calculation further. The time instant at which this happens
is dubbed the threshold time $t_\text{th}$. 
Depending on the variance $\sigma^2$ and on the magnetic field $h$ we are able to reach 
different values of $t_\text{th}$. For times $t>t_\text{th}$ the data from the DMRG 
calculations are not reliable so that we are not able to calculate the coherence 
times $T_2$ larger than $t_\text{th}$. This occurs especially for small variances $\sigma^2$
and/or small magnetic fields $h$. But we emphasize that the accumulated discarded weight is well below $10^{-5}$ for $t=T_2$ except for special cases.

Furthermore, we are limited by the required CPU time. Hence we do not 
investigate time scales $A_Q t > 40$.

\subsubsection{Dependence of the coherence time on the initial variance}

\begin{figure}[htb]
 \centering
 \includegraphics[width=\columnwidth]{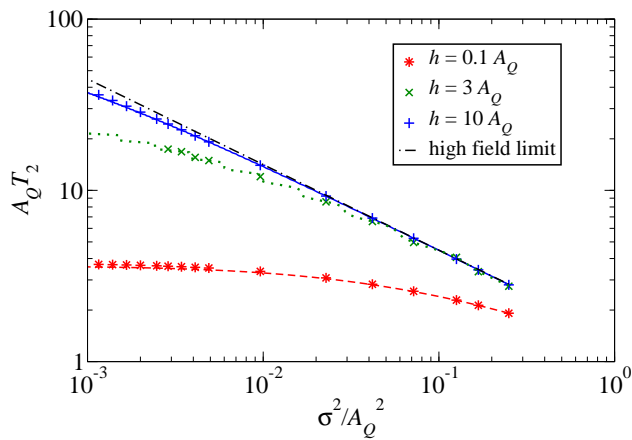}
 \caption{Coherence time $T_2$ for various external magnetic fields $h$ 
depending on the initial variance $\sigma^2$. The data points are calculated for $N = 49$ 
bath spins and nonuniform couplings with $x = 1$. The colored lines {(dashed, dotted and solid)} are calculated for 
$N = 999$ bath spins and uniform couplings. The black dash-dotted line is the curve for the 
effective high-field model in \eqref{eq:highfield_limit}.} 
 \label{fig:T2_sigma}
 \end{figure}

In Fig.\ \ref{fig:T2_sigma}, we plot $T_2(\sigma^2)$ for various magnetic fields $h$ 
depending on $\sigma^2$. In addition, the corresponding coherence time $T_2$ is depicted
in the high-field limit \eqref{eq:T2_infty}.
The  curves computed for uniform and  nonuniform couplings almost coincide. 
This observation agrees with previous works \cite{coish07,hackm14a}
 stating that the short-time dynamics depends 
hardly on the distribution of the coupling constants $A_k$.

For all magnetic fields $h$, the coherence time $T_2$ increases for decreasing variance 
$\sigma^2$ as expected. We observe, however, a qualitative difference between the 
curves for $h = 0.1A_Q$ and for $h = 3A_Q$ or $h = 10A_Q$. We discuss 
the low-field and high-field regimes separately.

In the low-field regime the slope of the coherence time $T_2(\sigma^2)$ falls quickly so that
$T_2$ does not grow strongly as $\sigma^2$ decreases. For $\sigma^2 < 0.0005A_Q$ the coherence time takes roughly 
 double the value of $T_2(\sigma^2 = A_Q^2/4)$ for the variance without any narrowing, i.e., 
$\gamma=0$. The fact that $T_2$ increases only weakly for lowered
variance can be easily explained by the flip-flop terms. 
These terms are not influenced by the narrowing of the spin bath and affect the central 
spin equally for any value of $\gamma$. Hence the decoherence of the central spin cannot 
be suppressed efficiently.

In the  regime of larger fields $h = 3 A_Q$ and $h = 10 A_Q$ we observe a different behavior in 
Fig.\ \ref{fig:T2_sigma}. The data matches the thermodynamic limit
\eqref{eq:highfield_limit} 
well for variances {$\sigma^2 \gtrapprox A_Q^2/50$}. For these values of the variance the 
coherence time $T_2$ is almost independent of the magnetic field $h$ as long as the 
field is still large enough, i.e., the system is still in the high-field regime.
The data deviates from the thermodynamic limit for lower values of $\sigma^2$. For 
$h = 10 A_Q$ the data agrees with the formula down to lower variances than 
for $h = 3 A_Q$. Nonetheless, even for $h = 10A_Q$, we clearly see deviations for 
$\sigma^2 < A_Q^2/100$.

The coherence time for finite moderate magnetic fields grows more slowly than it does
in the high-field limit. But $T_2(\sigma^2)$ continues to increase for 
decreasing $\sigma^2$ and it may diverge for $\sigma^2 \rightarrow 0$ even for 
finite fields $h$. This would imply that the absolute value of the correlation function in 
\eqref{eq:correlation} does not fall below $1/e$. We have not found any signature
of this scenario in the transverse spin dynamics perpendicular to a finite magnetic field. 
Indeed, even for vanishing initial Overhauser fluctuations the flip-flop terms
{cause the central spin to exchange its $z$-polarization} with the bath.
So, on one hand, $\sigma^2=0$ does not imply the absence of decoherence.
On the other hand, we recall that  rigorous arguments show that persisting 
correlations are generic in the CSM without magnetic field
if the distribution of couplings is normalizable \cite{uhrig14a,seife16}. 
This does not even require that $\sigma^2=0$ holds.

\begin{figure}[htb]
 \centering
 \includegraphics[width=\columnwidth]{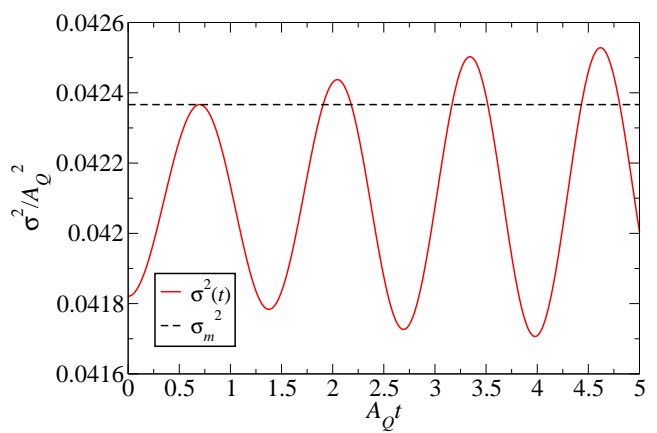}
 \caption{Time-dependent variance $\sigma^2(t)$ of the Overhauser field for 
nonuniform couplings with coupling spread $x = 1$ in 
\eqref{eq:inhomogeneous_couplings},  narrowing factor $\gamma = 100$, and
external magnetic field $h = 5A_Q$. In addition, we plot the value of the first 
maximum $\sigma^2_m$ of the time-dependent variance as a black dashed line.}
 \label{fig:sigma_dyn}
\end{figure}

In Fig.\ \ref{fig:sigma_dyn}, we plot the time-dependent variance $\sigma^2(t)$ for 
the spread parameter $x = 1$ and the narrowing factor $\gamma = 100$ to illustrate the 
temporal evolution of $\sigma^2(t)$.  Clearly, $\sigma^2(t)$ oscillates which is mainly
induced by the external magnetic field $h$. In addition, there is a trend to increase.
We use the first maximum $\sigma^2_m(t_m)$ of 
the oscillations to quantify the temporal evolution in the limit of  vanishing 
initial variance $\sigma^2$, i.e., $\gamma \rightarrow \infty$.

\begin{figure}[htb]
 \centering
 \includegraphics[width=\columnwidth]{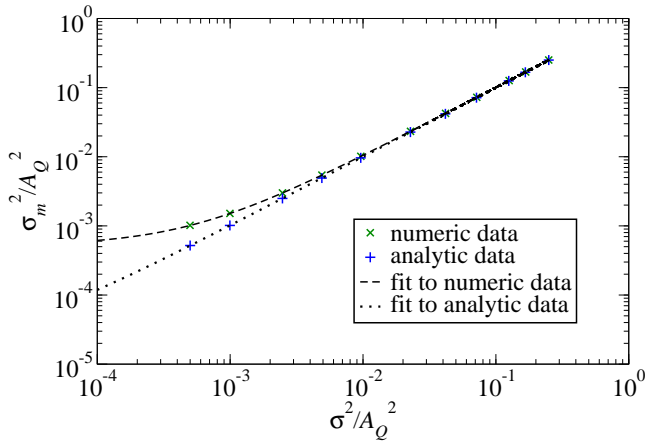}
 \caption{First maximum of the dynamic variance $\sigma^2(t)$ at time $t_m$, 
see Fig.\ \ref{fig:sigma_dyn}, depending on the initial variance $\sigma^2$ for external 
magnetic field $h = 5$ $A_Q$. {The green crosses are calculated for $N = 49$ bath spins and 
nonuniform couplings with $x = 1$. The dashed line is the corresponding quadratic fit. 
The blue crosses are calculated for $N = 999$ bath spins and 
uniform couplings. The dotted line is the corresponding linear fit. 
Both fits yield finite values of $\sigma^2_m$ for $\sigma^2\rightarrow0$.}}
 \label{fig:sigma_t}
\end{figure}

In Fig.\ \ref{fig:sigma_t}, we plot the value of the first maximum $\sigma^2_m$ of 
the time-dependent variance as a function of the initial variance $\sigma^2$ for 
 external magnetic field $h = 5 A_Q$. 
The curve decreases with decreasing $\sigma^2(0)$ for both uniform and nonuniform couplings. 
To analyze the behavior of $\sigma^2_m$  in the limit $\sigma^2\rightarrow 0$  
we fit the polynomials $f_1$ and $f_2$ which take the form
\begin{subequations}
\begin{align}
 f_1(\sigma^2) &= a_1 + b_1 \sigma^2 + c_1 \sigma^4 \approx \sigma^2_m(x=1), 
\\
 f_2(\sigma^2) &= a_2 + b_2 \sigma^2 \approx \sigma^2_m(x=0),
\end{align}
\end{subequations}
to the data of the nonuniform and the uniform system, respectively. The 
fits are used to extrapolate the first maximum for $\sigma^2 \rightarrow 0$ yielding
 $a_1$ and $a_2$. Quantitatively, these values depend on the external magnetic field. 
But the qualitative behavior for all finite magnetic fields $h$ is the same. 
Hence it is sufficient to discuss the effect of $\sigma^2 \rightarrow 0 $ for one 
choice $h = 5 A_Q$ of the magnetic field.

Both fits yield finite values 
\begin{subequations}
\begin{align}
 a_1 &= \left(5.15224 \cdot 10^{-4} \pm 3.192 \cdot 10^{-6}\right) A_Q^2 
\\
 a_2 &= \left(1.84859 \cdot10^{-5} \pm 2.211 \cdot 10^{-8} \right) A_Q^2
\end{align}
\end{subequations}
for $\sigma^2 = 0$. Thus, we find the remarkable fact that the first maximum 
$\sigma_m(t_m)$ is finite even if the initial variance vanishes. This is
shown in Fig.\ \ref{fig:sigma_t} for $h = 5 A_Q$. For different 
magnetic fields the numeric values change but stay finite so that the qualitative
finding remains the same. Note that this observation supports the above argument
that the flip-flop terms induce decoherence even if the initial variance vanishes.
One mechanism is that the variance is not constant and increases in the course
of time even if it was zero in the beginning.

Since the fluctuations of the Overhauser field are finite for $t>0$ the correlation in 
\eqref{eq:correlation} decreases in time even for $\sigma^2(t=0) \rightarrow 0$. 
Still the limit of a vanishing initial variance $\sigma^2$ yields the best possible 
reduction of decoherence. Thus we want to determine the longest possible coherence time.
To this end, we conceive an extrapolation scheme 
to calculate the absolute correlation function $\left|C(t)\right|$ for 
$\sigma^2 \rightarrow 0$. As pointed out in the previous subsection the limit 
$\sigma^2 \rightarrow 0$ leads easily to finite size artifacts for uniform
couplings. We can use even larger bath sizes $N$ to calculate $C(t)$ for 
uniform couplings. The results are very similar to the corresponding 
results for nonuniform couplings. Hence we restrict the analysis to the 
correlation for nonuniform couplings with spread parameter $x=1$.

\begin{figure}[htb]
 \centering
 \includegraphics[width=\columnwidth]{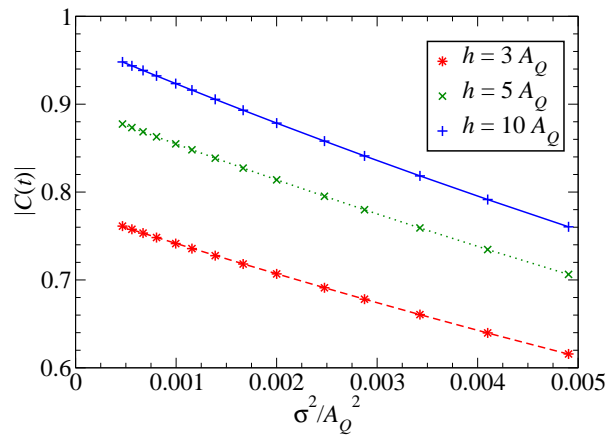}
 \caption{Absolute value of the correlation function $C(t)$ for $A_Q t = 10$ 
and its dependence on the variance $\sigma^2$ for three magnetic fields and
nonuniform couplings with $x = 1$ in \eqref{eq:inhomogeneous_couplings}. 
The colored lines {(dashed, dotted and solid)} are fits of the form \eqref{eq:corr_fit} to the data points.} 
 \label{fig:time_sigma}
 \end{figure}

From the data we determine the dependence of $\left | C\left(t \right) \right |$ 
on the variance $\sigma^2$ at fixed time $t$. 
In Fig.\ \ref{fig:time_sigma},
we plot $\left | C\left(t \right) \right |$ for three different fields $h$ depending 
on the initial variance $\sigma^2$. All three graphs can be approximated very well 
by the function
\begin{align}
 \left| C\left(t \right) \right | = C_0(t) \exp \left( - a(t) \sigma^2 \right) .
\label{eq:corr_fit}
\end{align}
Here, the parameters $C_0(t)$ and $a(t)$ are fitted for the fixed time $t = 10/A_Q$. The 
resulting fits are displayed in Fig.\ \ref{fig:time_sigma} as well.

For each time $t$ we determine the parameters $C_0(t)$ and $a(t)$ in 
\eqref{eq:corr_fit} by fits to the numeric data. 
Since we are interested in the limit $\sigma^2 \rightarrow 0$ the absolute 
value of the correlation $C(t)$ in this limit is given by $C_0(t)$. The prefactor 
$C_0(t)$ represents the correlation with the largest possible decoherence time $T_2$ for a 
given magnetic field $h$ because the limit $\sigma^2 \rightarrow 0$ yields the best 
possible reduction of fluctuations of the Overhauser field.

\begin{figure}[htb]
 \centering
 \includegraphics[width=\columnwidth]{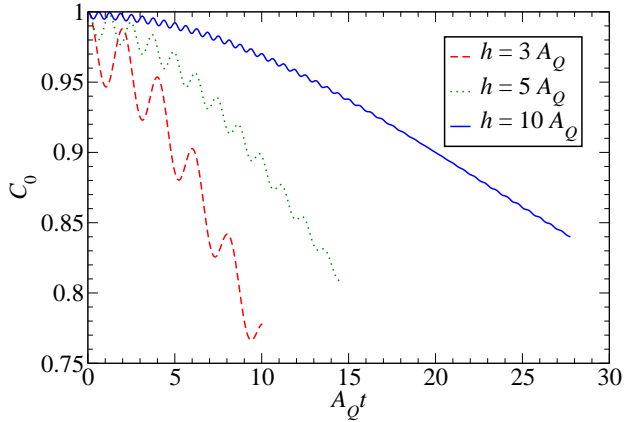}
 \caption{Fit parameter $C_0(t)$ in \eqref{eq:corr_fit} depending on time $t$ for 
three different magnetic fields $h$ and nonuniform couplings with $x = 1$ in
\eqref{eq:inhomogeneous_couplings}. The parameter $C_0(t)$ is equal to the absolute value 
of the correlation function $C(t)$ in the limit $\sigma^2 \rightarrow 0$. Hence the 
curves represent the correlation functions for optimally suppressed 
fluctuations in the spin bath and yield the maximum coherence time $T_2$.} 
 \label{fig:time_infty}
 \end{figure}

In Fig.\ \ref{fig:time_infty}, the resulting $C_0(t)$ is displayed
for three different magnetic fields $h$. 
The correlation $C_0(t)$ decreases in time, but it does so much {more slowly} than for finite 
$\sigma^2$. With increasing magnetic fields $h$ the correlation decreases {more and more slowly}. 
Hence one can achieve higher coherence times $T_2$ in this limit as can also be seen in 
Fig.\ \ref{fig:T2_sigma}. 

\subsubsection{Dependence on the external magnetic field}

In Figs.\ \ref{fig:T2_sigma} and \ref{fig:time_infty}, we observe clearly that the 
decoherence of the central spin is influenced by the applied magnetic field $h$. In both 
figures, the coherence time grows with increasing magnetic field $h$. 
The solution \eqref{eq:highfield_limit} for infinite magnetic field represents
the upper limit of the coherence time $T_2$. 
Here, we focus on more details as they are relevant in any experimental setup
which is described by the CSM.

\begin{figure}[htb]
 \centering
 \includegraphics[width=\columnwidth]{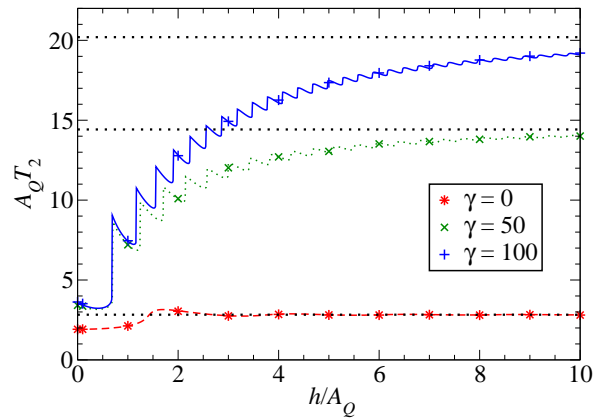}
 \caption{Coherence time $T_2$ as function of the external magnetic field $h$
for various narrowing factors $\gamma$. The data points are calculated for 
$N = 49$ bath spins and nonuniform couplings with $x = 1$ in 
\eqref{eq:inhomogeneous_couplings}. The colored lines {(dashed, dotted and solid)}
 are calculated for $N = 999$ 
bath spins and uniform couplings. The dotted black lines are the corresponding 
coherence times $T_{2,\infty}$ in \eqref{eq:T2_infty} for the three values of 
$\gamma$ in the high-field limit $h\rightarrow \infty$.}
 \label{fig:T2_h}
 \end{figure}

We plot the coherence time $T_2$ as function of $h$ 
for three narrowing factors $\gamma$ in Fig.\ \ref{fig:T2_h}. In addition,
we include the corresponding coherence time $T_{2,\infty}$ in the high-field limit 
from \eqref{eq:T2_infty} for the three values of $\gamma$. 
Remarkably, the curve for $\gamma = 0$ behaves qualitatively different. 
Hence we will discuss the cases $\gamma = 0$ and $\gamma > 0$ separately.

For $\gamma=0$, the coherence time $T_2$ depends hardly on the external magnetic field $h$. 
The curve overshoots the high-field limit $T_{2,\infty}$ slightly, but 
approaches  $T_{2,\infty}$ in \eqref{eq:T2_infty} quickly for higher fields $h$. 
This finding agrees with previous results \cite{stane14b} where the CSM was
analyzed as well without any narrowing. We point out that the coherence time $T_2$ in 
Ref.\ \onlinecite{stane14b} was determined by a Gaussian fit to a different 
correlation function and was thus not defined in precisely the same way as in the present work. 
Nonetheless, the results are qualitatively the same.  
Also  Ref.\ \onlinecite{hackm14a} showed  that for sufficiently large 
fields the decoherence of the system is independent of the magnetic field.

In contrast to the curve without narrowing,
 the two curves for $\gamma = 50$ and $\gamma = 100$ show an overall 
increase of the coherence time $T_2$ with growing magnetic field $h$ besides the
sawtooth features which we explained at the beginning of this subsection.
The coherence time $T_2$ does not increase without bound, but it converges to the value
$T_{2,\infty}$. The physics is easily explained. 
By increasing the magnetic field the influence of the flip-flop terms 
of the Hamiltonian in \eqref{eq:Gaudin} is decreased. Then, the fluctuations of the Overhauser 
field become the main source of decoherence.  Since we narrow 
the distribution of the Overhauser field, i.e., we reduce the
detrimental fluctuations the coherence time can be increased.
This works efficiently for high magnetic fields because at low fields the
flip-flop mechanism is still at work inducing decoherence which is 
unrelated to the initial variance $\sigma^2$.

This observation allows us also to establish a connection to the 
common distinction between longitudinal relaxation and transverse
relaxation or dephasing. Both processes together yield decoherence.
But a sizable magnetic field is required to separate them clearly.
For large field the longitudinal relaxation is reduced and suppressed
in the high-field limit because  transitions
via the flip-flop terms are its prerequisite. Then, decoherence reduces to pure dephasing. 
For moderate magnetic fields, however, dephasing and longitudinal relaxation
are both at work.

For a quantitative analysis, we inspect the convergence of the coherence time $T_2$ 
for a finite field $h$ to the infinite-field limit $T_{2,\infty}$.
{This process can be assessed} by the relative deviation
\begin{align}
\Delta T_2 = \left | \frac{T_2}{T_{2,\infty}} - 1 \right | 
\label{eq:T2_relDiff}.
\end{align}
For small fields the deviation is quite large. In this regime the 
flip-flop terms in the Hamiltonian are still active in the dynamics of 
the central spin. Upon increasing field $h$ the deviation $\Delta T_2$ 
decreases and fulfills
\begin{align}
 \Delta T_2 \propto h^{-2} \quad \text{for } h > 5 A_Q.
\end{align}
 In this regime, the dynamics of the central spin is dominated by the
$\hat B_z\hat S_z$ interaction between the 
central spin and the Overhauser field. Hence the fluctuations of the 
Overhauser field dominate the dephasing. By narrowing its
initial distribution one can efficiently increase 
the coherence time $T_2$ in the high-field limit $h \gg 1 A_Q$.

\section{Conclusion}
\label{sec:conclusion}

The central spin model describes a wide range of decoherence phenomena
due to spin baths, for instance in quantum dots. One important proposal
to reduce decoherence is to suppress the fluctuations in the bath.
The key quantity is the Overhauser field, i.e., the sum of the effects
of all bath spins on the central spin.
Thus, the goal of our study was to investigate the effect of narrowed
distributions of the Overhauser field. To this end, we used and extended two
state-of-the-art techniques for the central spin model. 
We calculated the dynamics of the central spin for 
large spin baths by numeric DMRG for nonuniform distribution
of couplings and by an analytic solution for uniform couplings.

By introducing the narrowing factor $\gamma$ we  adjusted the initial variance 
$\sigma^2$ of the Overhauser field. The narrowing factor $\gamma$ and $\sigma^2$ 
are related by a one-to-one mapping, i.e., for given values of the spread parameter $x$ 
and the bath size $N$ the variance is determined by $\gamma$. Our study revealed 
that the initial variance depends only slightly on the spread parameter $x$ and on 
the bath  size $N$ for a wide range of parameters.

We showed that generally the coherence time $T_2$ of the central spin model can be increased 
substantially by narrowing the distribution of Overhauser field. 
There is, however, an important restriction to this statement.
The coherence time does not grow upon decreasing the variance $\sigma^2$
of the Overhauser field independent of the applied magnetic field. Without field, the
narrowing is almost pointless. Our results extend previous findings
\cite{stepa06,onur14} dealing with high magnetic fields. 
Only in this limit, the relation $T_2\propto 1/\sigma$ is valid.

For low magnetic fields the coherence time $T_2(\sigma)$ is  limited roughly by twice
the  value $T_2(\sigma_0)$ without any narrowing. The 
dynamics is driven by the flip-flop terms which are unchanged by 
narrowing the distribution of the Overhauser field. 
Hence the coherence of the central spin decays quite 
fast even for very small values of $\sigma^2$. Therefore, the coherence time is almost 
independent on $\sigma^2$ for narrowly distributed Overhauser fields.

With increasing magnetic field the coherence time increases because the flip-flop 
terms are suppressed more and more. Nonetheless, the coherence time increases more slowly 
than the inverse of $\sigma$. In addition, we showed that the central spin decoheres 
even in the limit of vanishing initial variance $\sigma^2$ because
the flip-flop terms are still active.

In the high-field regime, the system can be approximated by an effective Hamiltonian 
containing no flip-flop terms{\cite{witze06,koppe07}.} 
{In hole-doped systems, the flip-flop terms are reduced from the beginning
\cite{fisch08a,teste09,hackm14a}. 
Without the flip-flop terms,} the coherence time is 
indeed inversely  proportional to $\sigma$. 
Hence $T_2$ can be increased arbitrarily 
by decreasing $\sigma$. But the coherence time is bounded {for the cases where  
flip-flop terms are present}.

The initial variance $\sigma^2$ of the Overhauser field determines the 
maximum coherence time of the central spin system. This can be achieved for 
optimum conditions, i.e., for very large magnetic field applied to the central spin. 
The role of the external magnetic field is to suppress the
effect of the flip-flop terms. 

Further research in this field, for instance the investigation of
other distributions of the couplings or the effect of protocols
of dynamic decoupling, is certainly called for. Our study 
has illustrated that time dependent density-matrix renormalization is
a very useful numeric tool to this end.

\section{Acknowledgment}

We thank the Deutsche Forschungsgemeinsschaft for financial support in project 
UH90/9-1 and in the ICRC TRR~160 together with the 
Russian Foundation of Basic Research.

\appendix

\section{Time evolution for uniform couplings}
\label{app:time_evolution}

For uniform couplings in the CSM \eqref{eq:Gaudin} we can 
calculate the time evolution analytically. We have to calculate the time evolution 
for states of the form 
$\left | \uparrow \right \rangle \otimes \left | j,m \right \rangle$ and $\left |
\downarrow \right \rangle \otimes \left | j,m \right \rangle$. As argued in 
Sec.\ \ref{sec:methods_analytic} the Hamiltonian in \eqref{eq:Gaudin} is block diagonal 
for uniform couplings, thus, the time evolution can be calculated fairly easily
because the blocks are of dimension two only. 
 For the following calculation, we shift the Hamiltonian by $A_Q/\sqrt{N}4$
leading to
\begin{align}
 \hat{H}^\prime = \frac{A_Q}{\sqrt{N}} \hat{\mathbf{I}} \hat{\mathbf{S}} - h \hat{S}_z + 
\frac{A_Q}{4\sqrt{N}}.
\end{align}
This shift has no influence on expectation values of the system because
it cancels out in the traces.

For the two states $\left|1\right \rangle = \left | \uparrow \right \rangle \left 
| j,m \right \rangle$ and $\left|2\right \rangle = \left | \downarrow \right \rangle \left 
| j,m+1 \right \rangle$ with $\left|m\right| <j$ the effect of the Hamiltonian is
\begin{subequations}
\begin{align}
 \hat{H}^\prime \left | 1 \right \rangle &= a_m \left | 1 \right \rangle + b_{j,m} 
\left | 2 \right \rangle, 
\\
 \hat{H}^\prime \left | 2 \right \rangle &= -a_m \left | 2 \right \rangle + b_{j,m} 
\left | 1 \right \rangle,
\end{align}
\end{subequations}
where we use the abbreviations
\begin{subequations}
\begin{align}
a_{m} &= \frac{\left(2m + 1\right)}{4} \frac{A_Q}{\sqrt{N}} - h/2, 
\\
b_{j,m} &= \frac{A_Q}{2\sqrt{N}}\sqrt{\left(j - m \right) \left(j + m + 1\right)}.
\end{align}
\end{subequations}
Diagonalizing the block matrices yields the eigenvalues
\begin{align}
 \omega^{\pm}_{j,m} = \pm \sqrt{a_{m}^2 + b_{j,m}^2} 
\label{eq:omega}
\end{align}
 of $\hat{H}^\prime$. With them the time evolution is given by
\begin{subequations}
\begin{align}
 \left | 1 \right \rangle(t) = \left(c_{j,m} - \mathfrak{i} s_{j,m}\right) 
\left | 1 \right \rangle - \mathfrak{i} \tilde{s}_{j,m} \left | 2 \right \rangle, 
\\
 \left | 2 \right \rangle(t) = \left(c_{j,m} + \mathfrak{i} s_{j,m}\right) 
\left | 2 \right \rangle - \mathfrak{i} \tilde{s}_{j,m} \left | 1 \right \rangle,  
\end{align}
\end{subequations}
where we employ the abbreviations
\begin{subequations}
\begin{align}
c_{j,m} &= \cos\left(\omega_{j,m} t\right), 
\\
s_{j,m} &= \sin\left(\omega_{j,m} t\right) \frac{a_{m}}{\omega_{j,m}},
\\
\tilde{s}_{j,m} &= \sin\left(\omega_{j,m} t\right) \frac{b_{j,m}}{\omega_{j,m}}.
\end{align}
\end{subequations}

Two states are left out so far, namely 
$\left | + \right \rangle \coloneqq 
\left | \uparrow \right \rangle \left | j,j \right \rangle$ and 
$\left | - \right \rangle \coloneqq \left | \downarrow \right \rangle 
\left | j,-j \right \rangle$. They are eigenstates of $\hat{H}^\prime$ and 
can be evolved in time directly yielding
\begin{align}
\left | \pm \right \rangle(t) = 
\exp\left[-\mathfrak{i}\left(\frac{A_Q(2j+1)}{4\sqrt{N}}\pm\frac{h}{2}\right)t\right] 
\left | \pm \right \rangle.
\end{align}

The last step is to calculate the expectation values $\sigma^2(t)$ in 
\eqref{eq:sigma_result}  and $C(t)$ in \eqref{eq:correlation}. 
For a lighter notation we define the weighted sums
\begin{align}
 \overline{f_{j,m}} \coloneqq \sum_{j,m} g_{j,m}(\gamma) f_{j,m}.
\end{align}
The weight $g_{j,m}$ is defined in \eqref{eq:g_jm}. {The maximum of $j$ is $j_\text{max} = N/2$ while its minimum is $j_\text{min} = 0$ or $j_\text{min} = 1/2$ for an even or odd $N$, respectively. 
The quantum number $m$ ranges from $-j$ to $j$.} and the range of the indices 
$j$ and $m$ is given in Sec.\ \ref{sec:methods_analytic}.

For the variance $\sigma^2(\gamma)$ we need to calculate two expectation values. 
With the help of the time evolved states we obtain
\begin{subequations}
\begin{align}
 \left \langle \hat{B}_z \right \rangle &= \frac{A_Q}{\sqrt{N}} \left(\overline{m} + 
\overline{\tilde{s}_{j,m}^2}\right)
\\
 \left \langle \hat{B}_z^2 \right \rangle &= \frac{A_Q^2}{N} \left(\overline{m^2} + 
\overline{\left(2m+1\right)\tilde{s}_{j,m}^2} \right)
\end{align}
\end{subequations}
by taking the traces.
Since the narrowing of the distribution of the Overhauser field is symmetric with respect
to the quantum number $m$, the weighted sum $\overline{m}$ vanishes. 
For the initial variance $\sigma^2$ in \eqref{eq:sigma_initial} we obtain the simpler result
\begin{align}
 \sigma^2 = \frac{A_Q^2}{N} \overline{m^2}
\end{align}
because all terms proportional to $\sin\left(\omega_{j,m} t\right)$ vanish. Thus, we dispose
of the expressions to calculate the initial variance $\sigma^2$ and the time dependent 
variance $\sigma^2(t)$.

To find the correlation function $C(t)$, we need to compute only one expectation value. 
Using the given time evolution of states we obtain the solutions for the real  and 
imaginary parts of $C(t)$ 
\begin{subequations}
\begin{align}
 \Re{\left[C(t)\right]} &= \overline{ c_{j,m} c_{j,m-1}} - \overline{s_{j,m} s_{j,m-1}}, \label{eq:correlation_analyticA}
\\
 \Im{\left[C(t)\right]} &= \overline{c_{j,m} s_{j,m-1}} + \overline{s_{j,m} c_{j,m-1}}. \label{eq:correlation_analyticB}
 \end{align}
\end{subequations}

\section{Computing the narrowed state by DMRG} 
\label{app:construct_gamma}

In Sec.\ \ref{sec:methods_DMRG} we discussed how one can realize
the narrowed Overhauser fields by DMRG. Here we present further details.
The task is to construct the narrowed state $\left | \gamma \right \rangle$ 
by evaluating
\begin{align}
 \left | \gamma \right \rangle = \exp\left( - \frac{\gamma}{2} {\hat{B}}_z^2/{A_Q^2} \right) 
\left | \psi \right \rangle = \hat{M} \left | \psi \right \rangle. 
\label{eq:target}
\end{align}
The state $\left | \psi \right \rangle$ is the purified state as introduced previously in Ref.\ \onlinecite{stane13a}.

We can easily apply the exponential operator $\hat{M}$ in \eqref{eq:target} by expressing the purified state $\left | \psi \right \rangle$ in the eigenbasis of the Overhauser field $\hat{B}_z$, see below. In  DMRG, the purified state $\left | \psi \right \rangle$ is approximated
by a state in the product basis of the system block $S$ 
and the environment block $E$ 
after each DMRG step of a sweep, including the truncation of the basis.
In the following, vectors of the system block will be denoted by the subscript 
$S$ and vectors of the environment block by the subscript $E$. For more details regarding the 
steps and sweeps the reader is referred to Ref.\ \onlinecite{stane13a}.

After a step of the sweep, we can express the state 
$\left | \psi \right \rangle$ in the truncated product basis of system and environment by
\begin{align}
 \left | \psi \right \rangle = \sum_{i,j} \Psi_{ij} \left | i \right \rangle_S 
\left | j \right \rangle_E
\end{align}
with the coefficients $\Psi_{ij}$ and the vectors $\left | i \right \rangle_S$ and 
$\left | j \right \rangle_E$. In the DMRG algorithm we express the narrowed
state also as the purified state in the same basis, i.e., we use
\begin{align}
 \left | \gamma \right \rangle = \sum_{i,j} \Gamma_{ij} 
\left | i \right \rangle_S \left | j \right \rangle_E.
\end{align}
In order to calculate the coefficients $\Gamma_{ij}$, one  applies the exponential function 
$\hat{M}$ in \eqref{eq:target} to the purified state $\left | \psi \right \rangle$.

Note that the Overhauser field naturally splits 
into the part $\hat{B}_{z,S}$ from the 
system block and $\hat{B}_{z,E}$ from the environment block so that $\hat{B}_{z}=
\hat{B}_{z,S} + \hat{B}_{z,E}$. We denote the eigenvectors of $\hat{B}_{z,S}$
by  $\left | a \right \rangle_S$ with eigenvalues $\lambda_a$ and those of $\hat{B}_{z,E}$ by 
$\left | b \right \rangle_E$ with eigenvalues $\mu_b$. Then, 
the action of $\hat{M}$ amounts to
\begin{align}
 \hat{M} \left | a \right \rangle \left | b \right \rangle = \exp\left[-\frac{\gamma}{2} 
\left(\lambda_a + \mu_b \right)^2/A_Q^2 \right] \left | a \right \rangle_S 
\left | b \right \rangle_E .
\end{align}

In order to re-express a purified state from a DMRG step
in terms {of} the eigenbasis of $\hat{B}_{z,S}$ and $\hat{B}_{z,E}$
we perform the following transformation.
Let us assume that the state is first given in the truncated basis resulting from 
the DMRG step denoted by $\left | c \right \rangle_S$ and 
$\left | d \right \rangle_E$. The eigenbasis of $\hat{B}_{z,S}$ and 
$\hat{B}_{z,E}$ is denoted by $\left | a \right \rangle_S$ and 
$\left | b \right \rangle_E$. Then, the {desired} coefficients
of $\left | \psi \right \rangle$ are computed according to 
\begin{align}
 \Psi^\prime_{ab} = \sum_{c,d} \leftidx{_S}{\left \langle a \right. 
\left | c \right \rangle}{_S} \leftidx{_E}{\left \langle b \right. 
\left | d \right \rangle}{_E} \Psi_{cd} .
\end{align}
The exponential operator $\hat{M}$ can be straightforwardly applied 
to the state in the eigenbasis of $\hat{B}_{z,S}$ and $\hat{B}_{z,E}$.
This state is transformed back to the original truncated basis yielding the 
coefficients
\begin{align}
 \Gamma_{ij} = \sum_{a,b} \leftidx{_S}{\left \langle i \right. 
\left | a \right \rangle}{_S} \leftidx{_E}{\left \langle j \right. 
\left | b \right \rangle}{_E} M_{ab} \Psi^\prime_{ab}.
\end{align}
These coefficients provide the target state $\left | \gamma \right \rangle$ 
we need for the next DMRG step.

\section{Choice of weights} 
\label{app:weight}

In Sec.\ \ref{sec:methods_DMRG}, we described that the narrowed state 
$\left | \gamma \right \rangle$ can be constructed more accurately by 
adding two additional density matrices
\begin{subequations}
\begin{align}
 \hat{\rho}_1 &= \mathrm{Tr}_E\left[ \hat{B}_z 
\left | \psi \right \rangle \left \langle \psi \right | \hat{B}_z \right], 
\\
 \hat{\rho}_2 &= \mathrm{Tr}_E\left[ \hat{B}_z 
\left | \gamma \right \rangle \left \langle \gamma \right | \hat{B}_z \right ] 
\end{align}
\end{subequations}
to the reduced density matrix $\hat{\rho}^\prime_S$ in \eqref{eq:red_density_prime}.
The state $\left | \psi \right \rangle$ is the initial purified state without
narrowing.
The weighted sum 
\begin{align}
\hat{\rho}_{S} = w_0 \hat{\rho}^\prime_S + w_1 \hat{\rho}_1 + w_2 \hat{\rho}_2
\end{align}
is used to define the basis truncation in the DMRG step.
Clearly, $\hat{\rho}_{S}$ depends on the choice of the
normalized weights $w_0$, $w_1$, and $w_2$. 

To analyze the influence of the weights we define the relative error
of the partition function $Z$
\begin{align}
 \Delta Z(w_1,w_2) = \left | \frac{Z_\text{n}(w_1,w_2)}{Z_\text{a}} - 1 
\right |
\end{align}
 between the data calculated by DMRG and the analytic solution, 
see \eqref{eq:partition_func_DMRG} and \eqref{eq:partition_func_analytic}. 
We fix the parameter $\gamma = 50$ and the bath size $N = 49$, but we checked that 
for other sets of $\gamma$ and $N$ the error $\Delta Z$ displays the same 
qualitative behavior. Hence we restrict ourselves to this exemplary set.

\begin{figure}[htb]
 \centering
 \includegraphics[width=\columnwidth]{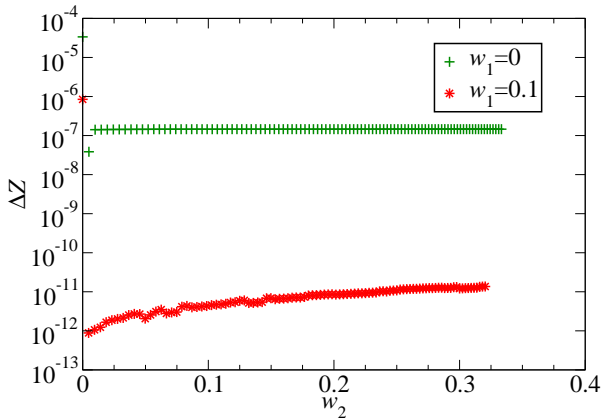}
 \caption{Relative error $\Delta Z(w_2)$ for two values of the weight $w_1$.}
 \label{fig:weight}
\end{figure}

In Fig.\ \ref{fig:weight}, $\Delta Z(w_2)$ is displayed for two values of $w_1$. 
Since the sum of the weights is unity the value of $w_0$ is implicitly
determined for the data points.  
The plot shows that for $w_1 = 0$ or $w_2 = 0$ the error is relatively large. The 
green data (crosses) is calculated for $w_1 = 0$; the relative error lies at about 
 $10^{-7}$ for most values of $w_2$. 

The red data (asterisks) lies at about $10^{-11}$ for $w_1 = 0.1$ and almost all 
values of $w_2$. For different values $w_1 > 0$ the corresponding error $\Delta Z$ 
is of the same order of magnitude. We conclude that one can choose the weights relatively
freely to obtain reliable results. But the total exclusion of one of 
the extra density matrices has to be avoided.
If this rule of thumb is complied with, the 
additional error introduced by the time evolution dominates.
{This case was studied elaborately} in Ref.\ \onlinecite{stane13a}.

\section{Accuracy check for the DMRG data}
\label{app:accuracy}

Here we present a more detailed analysis of the accuracy of the DMRG calculations. 
The analytic solutions for uniform couplings serve as benchmarks for this analysis.

During the build-up of the narrowed state $\left | \gamma \right \rangle$ we can 
easily calculate the partition function $Z_\text{n}$ 
according to \eqref{eq:partition_func_DMRG}. 
This result is compared to the analytic solution in 
\eqref{eq:partition_func_analytic}. 
For quantitative comparison, we define the relative error 
\begin{align}
 \Delta Z(\gamma,N) = \left | \frac{Z_\text{n}(\gamma,N)}{Z_\text{a}(\gamma,N)} - 1 
\right |
\end{align}
of the partition function $Z$ between the data calculated by DMRG and the 
analytic data, see \eqref{eq:partition_func_DMRG} and \eqref{eq:partition_func_analytic}. 
The error depends on the narrowing factor $\gamma$ as well as on the bath size $N$.

\begin{figure}[htb]
 \centering
 \includegraphics[width=\columnwidth]{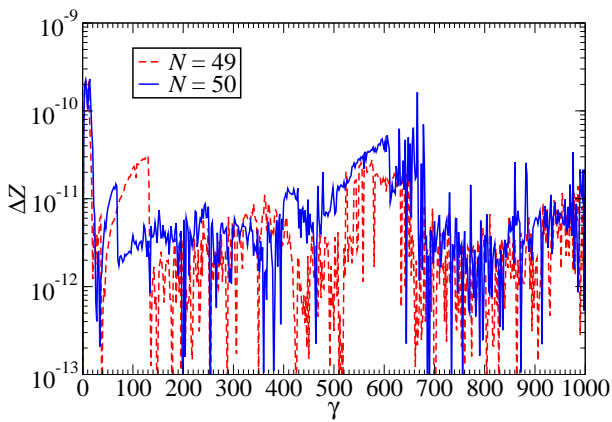}
 \caption{Relative error $\Delta Z(\gamma)$ as function of $\gamma$.}
 \label{fig:Z_gamma}
 \end{figure}

In Fig.\ \ref{fig:Z_gamma}, we show $\Delta Z(\gamma)$ for $N = 49$ and $N = 50$. 
Since the partition function behaves quite differently for odd and even values of $N$ 
data for both cases is included. For all values of $\gamma$, the 
error $\Delta Z$ is below $10^{-9}$ and for most values even below $10^{-11}$. We cannot 
identify a clear dependence of $\Delta Z$ on the narrowing factor $\gamma${, but} we observe
that the error is very small for all considered values of $\gamma$.

\begin{figure}[htb]
 \centering
 \includegraphics[width=\columnwidth]{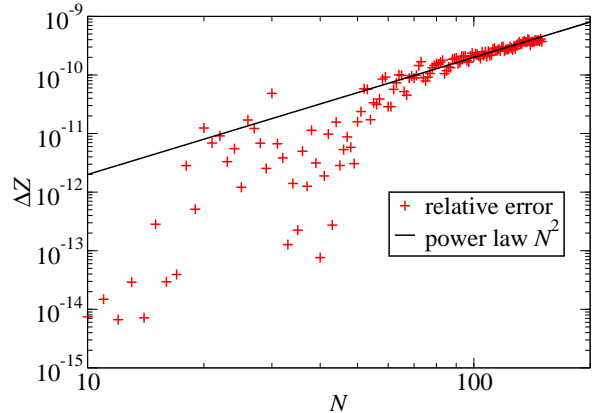}
 \caption{Relative error $\Delta Z(N)$ as function of $N$.}
 \label{fig:Z_N}
 \end{figure}

Figure\ \ref{fig:Z_N} shows the dependence of $\Delta Z(N)$ 
on the bath size $N$ for $\gamma = 200$. The relative error is again below 
$10^{-9}$ for all values of $N$. For larger bath sizes, we observe an increase of the
error roughly proportional to $N^2$. Since we consider mostly $N = 49$ in the
calculations presented in the main text the increase of the error is not an issue. 
Additionally,  the prefactor of the power law $N^2$ 
is of the order of $10^{-14}$. This means that one can treat large systems reliably
by DMRG if desired. 

In summary, the DMRG approach to create narrowed states is remarkably accurate. 
For time dependent quantities, the additional error due to the Trotter-Suzuki 
decomposition is dominating. This error has been comprehensively analyzed 
before \cite{stane13a}.

\section{Thermodynamic limit} 
\label{app:continuum_limit}

We study uniform and nonuniform couplings in systems with large bath size $N$. 
If we consider $N\to\infty$,  the central limit theorem helps us to 
express the tedious sums in the partition function $Z$ in \eqref{eq:partition_func_general} 
by a much simpler integral
\begin{align}
Z(\gamma) \approx Z_\infty(\gamma) \coloneqq \frac{2^{N}}{\sqrt{2\pi}\sigma_0} \int_{-\infty}^{\infty} 
e^{ - \frac{B_z^2}{2\sigma^2_0}} 
e^{ - \gamma \frac{B_z^2}{A_Q^2}} \mathrm{d}B_z
\end{align}
with the variance $\sigma^2_0 = A_Q^2/4$ in \eqref{eq:variance_zero} of the Overhauser field 
for  $\gamma = 0$ and the classical Overhauser field $B_z$, see Ref.\ 
\onlinecite{merku02} for more details regarding this limit. 
The partition function $Z_\infty$ denotes this thermodynamic limit.

Since the variance 
$\sigma^2_0$ is independent on the precise distribution of the couplings the result 
$Z_\infty$ is valid for uniform  and nonuniform couplings. The limit $Z_\infty$ 
can be calculated analytically yielding
\begin{align}
 Z_\infty(\gamma) = \frac{2^{N+1}}{\sqrt{4 + 2\gamma}}.
\end{align}
The corresponding variance $\sigma^2_\infty(\gamma)$ of the Overhauser field 
results from \eqref{eq:sigma_derivative} and we find
\begin{align}
\sigma^2_\infty(\gamma) = \frac{A_Q^2}{4 + 2 \gamma}. 
\label{eq:sigma_app}
\end{align}
In this way, we can describe the initial variance of the Overhauser field in 
the thermodynamic limit. By comparing \eqref{eq:sigma_app} with numeric results we 
investigated the finite size effects of the system for various coupling 
spreads $x$ and bath sizes $N$.

\section{High-field limit} 
\label{app:high_field_limit}

For an infinitely large external magnetic field an analytic solution 
for the correlation function \eqref{eq:correlation} of the central spin exists. The 
derivation is similar to the calculation in Refs.\  \onlinecite{merku02,cywin11}, but
we extend the previous result to narrowed distributions of Overhauser fields.

In the high-field limit $h \gg A_Q$ the 
Hamiltonian $\hat{H}$ in \eqref{eq:Gaudin} can be approximated by an effective Hamiltonian $\hat{H}_\text{eff}$ of the form
\begin{align}
 \hat{H}_\text{eff} = \hat{B}_z \hat{S}_z - h \hat{S}_z,
\end{align}
neglecting terms which change
$\hat{S}_z$ because their effect vanishes proportional to $1/h$. Hence the flip-flop terms vanish \cite{koppe07}. In addition, we consider the limit $N\rightarrow \infty$ to make use of the central  limit theorem. Combining both limits, the 
correlation function $C(t)$ in \eqref{eq:correlation} reduces to
\begin{align}
 C(t) \approx C_\infty(t) \coloneqq \frac{2^N e^{\mathfrak{i} h t}}{\sqrt{2\pi}\sigma_0 Z_\infty(\gamma)} I 
\end{align}
with the variance $\sigma^2_0 = A_Q^2/4$ in \eqref{eq:variance_zero} of the Overhauser field for $\gamma = 0$ and the partition function $Z_\infty(\gamma)$. The integral $I$ reads
\begin{align}
 {I \coloneqq \int_{-\infty}^{\infty} \exp \left( - \frac{B_z^2}{2\sigma^2_0} - 
\gamma \frac{B_z^2}{A_Q^2} - \mathfrak{i} B_z t\right) \mathrm{d}B_z }
\end{align}
with the classical Overhauser field $B_z$. It can be calculated analytically
\begin{align}
 C_\infty(t) = \exp\left(-\frac{\sigma^2_\infty(\gamma) t^2}{2}+\mathfrak{i} ht\right)
\end{align}
for the correlation function in the thermodynamic limit.
The initial variance $\sigma^2_\infty(\gamma)$ in \eqref{eq:sigma_app} is the 
variance in the thermodynamic limit.

Since we define the coherence time $T_2$ by $|C(T_2)| = 1/e$ we easily 
obtain the relation
\begin{align}
 T_{2,\infty} \coloneqq \frac{\sqrt{2}}{\sigma_\infty}
\end{align}
for the coherence time in the high-field limit.

\end{document}